\PassOptionsToPackage{numbers,super,sort&compress}{natbib}

\documentclass[
  aip,
  jcp,
  amsmath,amssymb,
  reprint,
  superscriptaddress
]{revtex4-1}

\usepackage{graphicx,amsmath,physics,xcolor} 
\usepackage{float}
\usepackage{achemso}

\usepackage{dcolumn}
\usepackage{bm}

\usepackage[utf8]{inputenc}
\usepackage[T1]{fontenc}
\usepackage{mathptmx}
\usepackage{etoolbox}

\makeatletter
\def\@email#1#2{%
 \endgroup
 \patchcmd{\titleblock@produce}
  {\frontmatter@RRAPformat}
  {\frontmatter@RRAPformat{\produce@RRAP{*#1\href{mailto:#2}{#2}}}\frontmatter@RRAPformat}
  {}{}
}%
\makeatother
\begin{document}

\preprint{AIP/123-QED}

\title[Cavity-Modified Nonequilibrium Fermi's Golden Rule Rate Coefficients from Cavity-Free Inputs]{Cavity-Modified Nonequilibrium Fermi's Golden Rule Rate Coefficients from Cavity-Free Inputs}
\author{Pouya Khazaei}
\author{Eitan Geva}%
 \email{eitan@umich.edu.}
\affiliation{ 
Department of Chemistry, University of Michigan, Ann Arbor, MI  48109, U.S.A.
}%
\date{\today}

\begin{abstract}
The Nonequilibrium Fermi’s Golden Rule (NE-FGR) provides a convenient theoretical framework for calculating the charge transfer rate between a photoexcited bright donor electronic state and a dark acceptor electronic state, when the nuclear degrees of freedom start out in a nonequilibrium initial state.
In this paper, we show that NE-FGR rates can be significantly modified by placing the molecular system inside an electromagnetic microcavity, even when the coupling with the cavity modes is weak. 
In this case, cavity-modified NE-FGR rates can also be estimated from the same inputs needed for calculating the cavity-free NE-FGR rates, 
thereby bypassing the need for an explicit simulation of the molecular system inside the cavity. We also introduce an approximate limit of the cavity-modified NE-FGR, which we denote cavity-modified Instantaneous Marcus Theory, since it is based on the same assumptions underlying Marcus theory. The utility of the proposed framework for calculating cavity-modified NE-FGR rates is demonstrated by applications to photo-induced charge transfer in the carotenoid-porphyrin-C$_{60}$ molecular triad dissolved in liquid tetrahydrofuran and the Garg-Onuchic-Ambegaokar model for charge transfer in the condensed phase.
\end{abstract}

\maketitle


\section{Introduction}

Charge transfer (CT) in donor-acceptor molecular systems represents an important class of chemical reactions.\cite{barbara96,chandler97,weiss99a,may00,nitzan06} 
They can occur spontaneously, as in redox reactions, or induced by electron injection, as in electrochemical devices, 
or by photoexcitation, as in photochemical and photovoltaic devices.\cite{liddell97,liddell02,bredas04,rizzi08,tian11,mishra09,feldt10,zhao12,lee13,lee14,lee14a,baiz10,strodel06,bredenbeck04,schmidtke04,xu94,ishizaki12,manna15,liang07}
In the case of photoexcitation from the ground electronic state to an excited bright donor state, 
the nuclear degrees of freedom (DOF) start out in a nonequilibrium initial state.
For example, in the case of impulsive excitation, the initial state of the nuclear DOF would correspond 
to equilibrium on the ground state potential energy surface (PES),
which is typically significantly different from that corresponding to equilibrium on the excited donor PES. 
   
A convenient way for describing the rate of photoinduced CT that starts out at the aforementioned nonequilibrium initial state is via the nonequilibrium Fermi's golden rule (NE-FGR).\cite{coalson94,cho95,izmaylov11,sun_nonequilibrium_2016,sun16d,hu20,brian21}
Similarly to equilibrium FGR (E-FGR),\cite{kubo55,jortner76,mikkelsen87,newton84,newton91,barbara96,nitzan06,zhao12,shi04b,nitzan06,sun16,sun16c,sun18,lee14,lee14a,schubert23} 
NE-FGR is based on the assumption of weak electronic coupling between donor and acceptor states (within the framework 
of second-order time-dependent perturbation theory). However, unlike E-FGR, NE-FGR allows for an arbitrary nonequilibrium initial state of the nuclear DOF and is able to capture the transient nonequilibrium dynamics prior to establishing E-FGR rate kinetics. This is manifested by the fact that NE-FGR describes the CT dynamics in terms of a {\em time-depepndent rate coefficient}, as opposed to a {\em time-independent rate constant}. 
It should also be noted that  NE-FGR is not limited to CT processes, and can be applied to other types of electronic transitions, 
such as electronic excitation energy transfer processes.\cite{jang02b}

NE-FGR is based on assuming that the dynamics of the nuclear DOF is governed by quantum mechanics. 
However, a fully quantum-mechanical calculation of NE-FGR rates is only possible when the donor and acceptor PESs are parabolic.  
For those cases, the fully quantum-mechanical expression for
the NE-FGR can be obtained in closed form and calculated in a feasible manner.\cite{kubo55,nitzan06,kestner74,jortner88,coalson94,leggett87,izmaylov11,endicott14,lee13,lee14,lee14a}
It should also be noted that the case of shifted, but otherwise identical, parabolic donor and acceptor PESs 
and constant donor-acceptor electronic coupling (the Condon approximation) corresponds to the canonical Marcus model for CT.\cite{marcus56a,marcus56b,marcus93}  
 Calculating NE-FGR CT rates for complex systems described by anharmonic force fields, for which a quantum-mechanically exact calcuation of the NE-FGR is not feasible, can be accomplished by utilizing semiclassical approximations.\cite{sun_nonequilibrium_2016,sun16d,hu20}

Recent experimental work has demonstrated that coupling
the molecular electronic and vibrational DOF
to the electromagnetic field modes of a microcavity can modify 
a wide range of chemical phenomena, including chemical reactivity and selectivity, photochemistry, catalysis, and electronic energy and charge transfer.\cite{andrew00,schwartz11,hutchison12,hutchison13,torma14,flick15,feist15,schachenmayer15,shalabney15,orgiu15,long15,ebbesen16,thomas16,zhong16,herrera16,casey16,sanvitto16,kowalewski16,kowalewski16a,flick17,zhong17,martinezmartinez17,fregoni18,saezblazquez18,flick19,galego19,lather19,schafer19,hoffmann19,hoffmann19a,lacombe19,mandal19,hoffmann20,flick20,gu20,mandal20,chowdhury21,saller21,basov21,saller_cavity-modified_2023,saller23a}
Those experimental advances call for the  development of theoretical models for estimating such cavity-induced modifications. 
Most recent theoretical studies of molecular matter in cavities assume that strong coupling between the molecular system and cavity modes is necessary for a significant cavity-induced effect to occur. 
The resulting strong-coupling models necessitate 
incorporating the cavity DOF into the model and explicitly 
simulating the molecular system inside the cavity.  
\cite{hoffmann19,hoffmann19a,hoffmann20,li20a,saller21,haugland21,chowdhury21,li21a,li21b,fischer22,wang22,li22,mandal22,mandal20,li20b}

However, even relatively weak coupling between molecular and cavity DOF can give rise to significant cavity-enabled effects. This is particularly true in the case of electronic energy and CT processes whose kinetics can be described in terms of FGR, which is based on treating the coupling between electronic states as a small perturbation.
Coupling between the electronic DOF and cavity modes can modify the coupling between electronic states when the molecular system is
placed inside a cavity. Treating this additional electronic coupling as a small perturbation can therefore still give rise to significant modifications of the corresponding FGR-based rates.

In a recent paper,\cite{saller_cavity-modified_2023} we pursued such an approach to show how cavity-modified E-FGR rate
constants can be estimated from inputs obtainable exclusively from a simulation of the cavity-free molecular system, thereby bypassing the need for an explicit simulation of
the molecular system inside the cavity. We also demonstrated the  usefulness of this framework in the case where the cavity-free kinetics is governed by Marcus theory, which corresponds to a semiclassical limit of E-FGR.

In this paper, we extend the work in Ref. \citenum{saller_cavity-modified_2023} to the case of the NE-FGR. We show that, similarly to the E-FGR case, the NE-FGR rate can be calculated from the same cavity-free inputs used for calculating the corresponding cavity-free NE-FGR rate for the same molecular system. This in turn allows us to bypass the need for an explicit simulation of the molecular
system inside the cavity in order to estimate cavity-induced modifications. We also introduce an approximation of the cavity-modified NE-FGR, which we denote
cavity-modified Instantaneous Marcus Theory (IMT), since it is based on the same assumptions underlying Marcus theory. 

The remainder of this paper is organized as follows. 
The theoretical framework for calculating cavity modified NE-FGR rates is presented and its asymptotic behavior analyzed in Sec. \ref{sec:theory}.
In Sec. \ref{sec:cpc60}, the proposed framework is used to calculate cavity-modified NE-FGR CT rates in the photoexcited carotenoid-porphyrin-C$_{60}$ (CPC$_{60}$) molecular triad dissolved in liquid tetrahydrofuran (THF) \cite{tong20, brian21,sun18} 
Another application to the benchmark
Garg-Onuchic-Ambegaokar (GOA) model for CT in the condensed phase\cite{garg85} is provided in Sec. \ref{sec:model}. 
A summary of the main results and concluding remarks are provided in Sec. \ref{sec:summary}.

\section{Theory}
\label{sec:theory}

\subsection{Preliminary considerations}
\label{subsec:prelim}

Adopting the same terminology as in Ref. \citenum{saller_cavity-modified_2023}, we consider a donor-acceptor molecular system that can undergo electronic CT, which is placed inside an electromagnetic cavity. The Hamiltonian of the system under consideration has the following form:
\begin{equation}	\hat{H}=\hat{H}^{np}_D\dyad{D}+\hat{H}^{np}_A\dyad{A}+\hat{V}^{np}_{DA}[\dyad{D}{A}+\dyad{A}{D}]~~.
\label{eq:overall_H}
\end{equation}
Here, $\hat{H}^{np}_j$ is the Hamiltonian of the nuclear (denoted by the superscript $n$) and photonic (denoted by the superscript $p$) DOF when the system is in the diabatic electronic state $\ket{j}$, where $j=D,A$ ($D$ and $A$ correspond to the donor and acceptor diabatic states, respectively): 
		\begin{equation}
			\hat{H}^{np}_j = \hat{H}^{n}_j+\hat{H}^{p}~~. \label{eq:ham}
		\end{equation}
It should be noted that $\hat{H}^{np}_j$ consists of a sum of a purely nuclear Hamiltonian, $\hat{H}^{n}_j$, which is electronic-state specific (i.e. dependent on $j$), and a purely photonic term, $\hat{H}^{p}$, which is the same regardless of the electronic state (i.e. independent of $j$). 
$\hat{V}^{np}_{DA}$ in the second term in Eq. \ref{eq:overall_H} is the coupling between the donor and acceptor states, which also consists of a sum of a purely nuclear and a purely photonic terms:  
		\begin{equation}
\hat{V}^{np}_{DA}=\hat{V}^{n}_{DA}+\hat{V}^{p}_{DA}~~. 
\label{eq:coup}
\end{equation}
The explicit forms of $\hat{H}^{n}_D$, $\hat{H}^{n}_A$, $\hat{H}^{p}$,
$\hat{V}^{n}_{DA}$ and $\hat{V}^{p}_{DA}$ are model-specific. Explicit examples for specific models will be presented below. However, for now, the derivation will remain general so as to be applicable to any form those operators may take. 

Focusing on the case where CT occurs between a photoexcited bright donor electronic state and a dark acceptor electronic state, the initial state of the overall system is assumed to be given by a density operator of the form:
\begin{equation}
\hat{\rho}(0)=\hat{\rho}_B(0)\otimes \dyad{D}~~.
\end{equation}
Here, $\hat{\rho}_B(0)$ is the initial state of the bath, which in our case consists of the nuclear and photonic DOF.
Further assuming that the initial state was created by impulsive photoexcitation from the ground state to an excited donor state, $\hat{\rho}_B(0)$ corresponds to the nuclear and photonic DOF being in thermal equilibrium with respect to the ground electronic state whose Hamiltonian is given by 
$\hat{H}^{np}_G = \hat{H}_G^{n} + \hat{H}^p$ (where $\hat{H}_G^{n}$ is assumed to be significantly different than $\hat{H}_D^{n}$):
\begin{equation}
\hat{\rho}_B(0)=
\hat{\rho}_n (0) \otimes \hat{\rho}_p (0)
=\frac{e^{-\beta\hat{H}^n_G}}{\Tr_n\left[e^{-\beta\hat{H}^n_G}\right]}\otimes\frac{e^{-\beta\hat{H}^p}}{\Tr_p\left[e^{-\beta\hat{H}^p}\right]}
~~.
\label{eq:initstate}
\end{equation}


\subsection{Cavity-modified population-only off-diagonal quantum master equation and nonequilibrium Fermi's golden rule rates}
\label{subsec:ne_fgr}
  
In the next step, we cast the Hamiltonian in Eq. \ref{eq:overall_H} in the form $\hat{H} = \hat{H}_0 + \hat{H}_I$, where 
the zero-order Hamiltonian, $\hat{H}_0$,
and the interaction Hamiltonian, 
$\hat{H}_I$, are give by:
\begin{equation}
\hat{H}_0 = \hat{H}^{np}_D\dyad{D}+\hat{H}^{np}_A\dyad{A}
\end{equation}
\begin{equation}
\hat{H}_I=\hat{V}^{np}_{DA}[\dyad{D}{A}+\dyad{A}{D}]
\end{equation}
 Treating $\hat{H}_I$ as a small perturbation within the framework of second-order perturbation theory and using projection operator techniques, one can derive the following equations of motion for the populations of the D and A states [${\sigma}_{DD}(t)$ and ${\sigma}_{AA}(t)$, respectively):\cite{lai_simulating_2021}
\begin{equation}
\frac{d}{dt}\sigma_{DD} (t)
=-\frac{d}{dt}\sigma_{AA} (t)
=-k_{D\rightarrow A}(t){\sigma}_{DD}(t)
+k_{A\rightarrow D}(t){\sigma}_{AA}(t) 
~~. 
\label{eq:p-tcl-OD-QME}
\end{equation}
Adopting the same terminology as in Ref. \citenum{lai_simulating_2021},
in what follows, we will refer to Eq. \ref {eq:p-tcl-OD-QME} as the {\em population-only time-convolutionless off-diagonal quantum master equation (p-tcl-OD-QME)}. 
Within the p-tcl-OD-QME, $k_{j\rightarrow k}(t)$ is the NE-FGR rate coefficient for population transfer from state $j$ to state $k$:
\cite{coalson94,cho95,izmaylov11,sun_nonequilibrium_2016,sun16d,hu20,lai_simulating_2021}
 \begin{align}
    & k_{j\rightarrow  k}(t) = \label{eq:nefgr} \\
    &\frac{2}{\hbar^2}\Re\int_0^t \dd \tau  \,  \Tr_B\left[e^{-\frac{i}{\hbar}\hat{H}^{np}_j t}\hat{\rho}_B(0) e^{\frac{i}{\hbar}\hat{H}^{np}_j t}\hat{V}^{np}_{jk} e^{-\frac{i}{\hbar}\hat{H}^{np}_k \tau}\hat{V}^{np}_{kj}e^{\frac{i}{\hbar}\hat{H}^{np}_j\tau}  \right] \notag~~.
\end{align}
Here, $\Tr_B[\cdot]=Tr_n Tr_p[\cdot]$ is the partial trace over the bath, which in our case corresponds to partially tracing over the nuclear and photonic Hilbert spaces. 

It should be noted that the quantum master equation in Eq. \ref{eq:p-tcl-OD-QME} relies on a certain choice of projection operator and the procedure used for deriving it. Other quantum master equations can be derived based on other choices of projection operators and following alternative procedures, which are not convolutionless and account for the dynamics of the coherences in addition to that of the populations (e.g., see Refs. \citenum{breuer02,trushechkin19,lai_simulating_2021} for a more detailed discussion). In what follows, we choose to focus on the p-tcl-OD-QME due to its relative simplicity compared to the other aforementioned variations and the fact that the dynamics is insensitive to the choice of equation for the parameter range characteristic of CT reactions.\cite{breuer02}

Inserting Eqs. \ref{eq:ham} and \ref{eq:coup} into Eq. \ref{eq:nefgr}, we obtain the following expression for the integrand on the R.H.S. of Eq. \ref{eq:nefgr}: 
\begin{align}
& \Tr_B\left[e^{-\frac{i}{\hbar}\hat{H}^{np}_j t}\hat{\rho}_B(0) e^{\frac{i}{\hbar}\hat{H}^{np}_j t}\hat{V}^{np}_{jk} e^{-\frac{i}{\hbar}\hat{H}^{np}_k \tau}\hat{V}^{np}_{kj}e^{\frac{i}{\hbar}\hat{H}^{np}_j\tau}  \right] \notag \\
      & = \Tr_n \left[ e^{-\frac{i}{\hbar} \hat{H}^{n}_j t} \hat{\rho}_n(0) e^{\frac{i}{\hbar} \hat{H}^{n}_j t} \hat{V}^{n}_{jk} e^{-\frac{i}{\hbar} \hat{H}^{n}_k \tau} \hat{V}^{n}_{kj} e^{\frac{i}{\hbar} \hat{H}^{n}_j \tau} \right] \notag\\
&\Tr_p \left[ e^{-\frac{i}{\hbar} \hat{H}^{p} t} \hat{\rho}_p(0) e^{\frac{i}{\hbar} \hat{H}^{p} t} e^{-\frac{i}{\hbar} \hat{H}^{p} \tau} e^{\frac{i}{\hbar} \hat{H}^{p} \tau} \right] \notag\\
&+ \Tr_n \left[ e^{-\frac{i}{\hbar} \hat{H}^{n}_j t} \hat{\rho}_n(0) e^{\frac{i}{\hbar} \hat{H}^{n}_j t} e^{-\frac{i}{\hbar} \hat{H}^{n}_k \tau} e^{\frac{i}{\hbar} \hat{H}^{n}_j \tau} \right] \notag\\
&\Tr_p \left[ e^{-\frac{i}{\hbar} \hat{H}^{p} t} \hat{\rho}_p(0) e^{\frac{i}{\hbar} \hat{H}^{p} t} \hat{V}^{p} e^{-\frac{i}{\hbar} \hat{H}^{p} \tau} \hat{V}^{p} e^{\frac{i}{\hbar} \hat{H}^{p} \tau} \right] \notag\\
&+ \Tr_n \left[ e^{-\frac{i}{\hbar} \hat{H}^{n}_j t} \hat{\rho}_n(0) e^{\frac{i}{\hbar} \hat{H}^{n}_j t} \hat{V}^{n}_{jk} e^{-\frac{i}{\hbar} \hat{H}^{n}_k \tau} e^{\frac{i}{\hbar} \hat{H}^{n}_j \tau} \right] \notag\\
&\Tr_p \left[ e^{-\frac{i}{\hbar} \hat{H}^{p} t} \hat{\rho}_p(0) e^{\frac{i}{\hbar} \hat{H}^{p} t} \hat{V}^{p} e^{-\frac{i}{\hbar} \hat{H}^{p} \tau} e^{\frac{i}{\hbar} \hat{H}^{p} \tau} \right] \notag \\
&+ \Tr_n \left[ e^{-\frac{i}{\hbar} \hat{H}^{n}_j t} \hat{\rho}_n(0) e^{\frac{i}{\hbar} \hat{H}^{n}_j t} e^{-\frac{i}{\hbar} \hat{H}^{n}_k \tau} \hat{V}^{n}_{kj} e^{\frac{i}{\hbar} \hat{H}^{n}_j \tau} \right] \notag \\
&\Tr_p \left[ e^{-\frac{i}{\hbar} \hat{H}^{p} t} \hat{\rho}_p(0) e^{\frac{i}{\hbar} \hat{H}^{p} t} e^{-\frac{i}{\hbar} \hat{H}^{p} \tau} \hat{V}^{p} e^{\frac{i}{\hbar} \hat{H}^{p} \tau} \right] \label{eq:bignefgr}
\end{align}
Importantly, according to Eq. \ref{eq:bignefgr}, cavity-modified NE-FGR rate coefficients can be obtained from purely nuclear and purely photonic time correlation functions as inputs and therefore do not necessitate an explicit simulation of the molecular system inside the cavity. Furthermore, as we will show below, the purely photonic time correlation functions can be obtained in closed form, while the purely nuclear correlation functions can be obtained from cavity-free models. In other words, {\em cavity-modified NE-FGR rate coefficients are obtainable from cavity-free inputs}.

To further demonstrate the implications of the aforementioned separability between nuclear and photonic time correlation functions, consider the case where the molecular system is coupled to a single cavity mode\cite{saller_cavity-modified_2023} (the extension to multiple modes is straightforward and leads to additive contributions to the NE-FGR rate from each of the modes\cite{saller23a}):
    \begin{equation}
    \hat{H}^p = \frac{\hat{p}_p^2}{2} +
    \frac{\omega_{p}^2\hat{q}_p^2}{2} ~~.
    \label{eq:p_Hamil}
    \end{equation}
Here, $\omega_{p}$ is the cavity mode's frequency, $\hat{q}_p$ is the photonic position operator and $\hat{p}_p$ is the photonic momentum operator. 
The electronic-photonic coupling
in this case is given by\cite{saller_cavity-modified_2023} 
\begin{equation}
\hat{V}^p_{DA}=\sqrt{2\hbar\omega_p}g_p\hat{q}_p=G\hat{q}_p~~,
\label{eq:p_couple}
\end{equation}
where,
\begin{equation}			g_p=\sqrt{\frac{\mu_{DA}^2\omega_p}{2\hbar\epsilon_0 V}}~~.
\end{equation}
Here, $\mu_{DA}$ is the donor-acceptor transition dipole moment and $V$ is the volume of the cavity.

Given Eqs. \ref{eq:p_Hamil} and \ref{eq:p_couple}, the photonic terms in Eq. \ref{eq:bignefgr} can be obtained in closed form:
\begin{align}
        & \Tr_p \left[ e^{-\frac{i}{\hbar} \hat{H}^{p} t} \hat{\rho}_p(0) e^{\frac{i}{\hbar} \hat{H}^{p} t} e^{-\frac{i}{\hbar} \hat{H}^{p} \tau} e^{\frac{i}{\hbar} \hat{H}^{p} \tau} \right] = 1~~, \notag \\
    				& \Tr_p \left[ e^{-\frac{i}{\hbar} \hat{H}^{p} t} \hat{\rho}_p(0) e^{\frac{i}{\hbar} \hat{H}^{p} t} \hat{V}^{p} e^{-\frac{i}{\hbar} \hat{H}^{p} \tau} e^{\frac{i}{\hbar} \hat{H}^{p} \tau} \right] =0~~, \notag \\
    				&\Tr_p \left[ e^{-\frac{i}{\hbar} \hat{H}^{p} t} \hat{\rho}_p(0) e^{\frac{i}{\hbar} \hat{H}^{p} t} e^{-\frac{i}{\hbar} \hat{H}^{p} \tau} \hat{V}^{p} e^{\frac{i}{\hbar} \hat{H}^{p} \tau} \right] = 0 ~~,\notag \\
				& \Tr_p \left[ e^{-\frac{i}{\hbar} \hat{H}^{p} t} \hat{\rho}_p(0) e^{\frac{i}{\hbar} \hat{H}^{p} t} \hat{V}^{p} e^{-\frac{i}{\hbar} \hat{H}^{p} \tau} \hat{V}^{p} e^{\frac{i}{\hbar} \hat{H}^{p} \tau} \right]  \notag \\
    &= \frac{\hbar G^2}{2 \omega_p} \cosh[\frac{\beta \hbar\omega_p}{2} - i\omega_p \tau ] \csch[\frac{\beta \hbar \omega_p}{2}]~~. 
    \label{eq:photeval}
\end{align}
Substituting the identities in Eq. \ref{eq:photeval} back into Eq. \ref{eq:bignefgr}, making the Condon approximation such that $\hat{V}^{n}_{jk} \rightarrow \Gamma$ (where $\Gamma$ is a constant, as opposed to a nuclear operator), and substituting the result back into Eq. \ref{eq:nefgr}, the {\em cavity modified NE-FGR rate coefficient} can be cast in the following form:
\begin{align}
    k_{j\rightarrow k}(t) &= \frac{2}{\hbar^2}\Re\int_0^t \dd \tau  \,  C^p(\tau) C^n_{jk}(t,\tau)~~, \label{eq:kprod}
\end{align}
where
\begin{align}
    C^n_{jk}(t,\tau)
    &=\Gamma^2 \Tr_n \left[ e^{-\frac{i}{\hbar} \hat{H}^{n}_j t} \hat{\rho}_n(0) e^{\frac{i}{\hbar} \hat{H}^{n}_j t} e^{-\frac{i}{\hbar} \hat{H}^{n}_k \tau}e^{\frac{i}{\hbar} \hat{H}^{n}_j \tau} \right] \label{eq:molcorrel}
\end{align}
and
\begin{equation}
			\begin{split}
     C^p(\tau) =& 1  + \frac{\hbar G^2}{2 \omega_p\Gamma^2} \cosh[\frac{\beta \hbar\omega_p}{2} - i \omega_p\tau] \csch[\frac{\beta \hbar \omega_p}{2}] \label{eq:photcorr}
\end{split}
\end{equation}
Here, $C^n_{jk}(t,\tau)$ captures the cavity-free nuclear dynamics and $C^p(\tau)$ captures the cavity effect. More specifically, decoupling the molecular system from the cavity mode, which corresponds to setting $G$ to $0$, implies that $C^p(\tau) \rightarrow 1$, such that $k_{j\rightarrow k}(t)$ reverts to the {\em cavity-free NE-FGR rate coefficient}:\cite{coalson94,cho95,izmaylov11,sun_nonequilibrium_2016,sun16d,hu20}
\begin{align}
    k_{j\rightarrow k}^m(t) &= \frac{2}{\hbar^2}\Re\int_0^t \dd \tau  \, C^n_{jk}(t,\tau)~~. 
    \label{eq:kprod11}
\end{align}
Here, the superscript $m$ refers to the fact that this is the NE-FGR rate coefficient for the cavity-free molecular system.


Further insight can be obtained by writing $C^p(\tau)$ in the following alternative, yet completely equivalent, form: 
\begin{align}
    C^p(\tau) =1 +\alpha \langle n_{eq} \rangle \left(e^{i \omega_p\tau} + e^{\beta \hbar\omega_p} e^{- i \omega_p\tau}\right)~~.
    \label{eq:expphotcorrel}
\end{align}
Here,  
\begin{align}
    \langle n_{eq}\rangle= \frac{1}{e^{\beta \hbar\omega_p}-1}
\end{align}
is the average number of photons in the cavity mode in thermal equilibrium and 
\begin{align}
    \alpha= \frac{G^2  \hbar }{2  \Gamma^2  \omega_p }~~.
\end{align}
Substituting Eq. \ref{eq:expphotcorrel} into Eq \ref{eq:kprod}, we obtain:
\begin{align}
    k_{j\rightarrow k}(t)&=k^m_{j\rightarrow k}(t) \label{eq:kshifted} 
    \\
    &+\alpha \langle n_{eq} \rangle \left[k^m_{j\rightarrow k}(t,\omega_p) + e^{\beta \hbar\omega_p} k^m_{j\rightarrow k}(t,-\omega_p) \right] \notag~~.
\end{align}
Here, $k^m_{j\rightarrow k}(t)$ is as in Eq. 
\ref{eq:kprod11}
and $k^m_{j\rightarrow k}(t,\pm \omega_p)$ is the NE-FGR rate coefficient for the dressed molecular system, where the PES of state $j$ is shifted by $\pm \hbar \omega_p$: 
\begin{widetext}
\begin{align}
    &k_{j\rightarrow k}^m(t,\pm \omega_p) = \frac{2}{\hbar^2}\Re\int_0^t \dd \tau  \, e^{ \pm i\omega_p\tau} C_n(t,\tau) 
    \label{eq:dress}
    \\
     &= \frac{2}{\hbar^2}\Re\int_0^t \dd \tau  \,  \Tr_n\left[e^{-\frac{i}{\hbar}(\hat{H}^{n}_j\pm\hbar\omega_p) t} \hat{\rho}_n(0) e^{\frac{i}{\hbar}(\hat{H}^{n}_j\pm\hbar\omega_p) t}\hat{V}^{n}_{jk}e^{-\frac{i}{\hbar}\hat{H}^{n}_k\tau}\hat{V}^{n}_{kj}e^{\frac{i}{\hbar}(\hat{H}^{n}_j\pm\hbar\omega_p) \tau}  \right] \notag
\end{align}
\end{widetext}


\subsection{Asymptotic behavior and cavity-modified instantaneous Marcus theory}
\label{subsec:asymptot}

Several limits of the cavity-modified NE-FGR rate coefficient in Eq. \ref{eq:kshifted} are noteworthy. One such limit corresponds to the case where the CT dynamics can be described in terms of the E-FGR rate constant. This corresponds to the case where the nuclear DOF start out at equilibrium on the parent state's PES, i.e. $\hat{\rho}_n (0) = \hat{\rho}_{n,j}^{eq} = \frac{e^{-\beta\hat{H}^n_j}}{\Tr_n\left[e^{-\beta\hat{H}^n_j}\right]}$, which implies that
\begin{equation}
C^n_{jk}(t,\tau) \rightarrow C^n_{jk}(\tau)
= \Gamma^2 \Tr_n \left[  \hat{\rho}_{n,j}^{eq} e^{-\frac{i}{\hbar} \hat{H}^{n}_k \tau}e^{\frac{i}{\hbar} \hat{H}^{n}_j \tau} \right]~~,
\end{equation}
such that:
\begin{align}
    k_{j\rightarrow k}(t) &\rightarrow \frac{2}{\hbar^2}\Re\int_0^t \dd \tau  \,  C^p(\tau) C^n_{jk}(\tau)~~, \label{eq:kprod1}
\end{align}
Further assuming that the lifetime of the correlation function $C^n_{jk}(\tau)$ is much shorter than the timescale of CT then allows one to extend the upper limit of the 
integral to infinity, such that it can be described by the {\em E-FGR rate constant} (as opposed to a time-dependent NE-FGR rate coefficient):
\begin{equation}
    k_{j\rightarrow k}^{eq}= 
    \frac{2}{\hbar^2}\Re\int_0^\infty \dd \tau  \,  C^p(\tau) C^n_{jk}(\tau)
\label{eq:cav_mod_efgr}
\end{equation}
A detailed analysis of the cavity-modified E-FGR rate constant in Eq. \ref{eq:cav_mod_efgr} has been recently reported in Ref. \citenum{saller_cavity-modified_2023}. 


Another noteworthy limit of Eq. \ref{eq:kshifted} is that of the Instantaneous Marcus Theory (IMT).\cite{tong20,brian21} 
To derive IMT, one starts with the short-time and high-temperature limit of the linearlized semiclassical (LSC) approximation for the cavity-free NE-FGR rate coefficient:\cite{hu20} 
\begin{align}
&k^{m}_{j\rightarrow k}(t)  \approx \frac{2}{\hbar^2}\Gamma^2 \Re \int_0^{t} d\tau \left\langle \exp \left[ 
\frac{i}{\hbar} U_{jk}({\bf R}_t ) \tau \right] \right\rangle~~.
\label{eq:LSC_st_ht}
\end{align}
Here, $U_{jk}({\bf R}_t) = V_j({\bf R}_t) - V_k({\bf R}_t)$, 
where $V_j({\bf R})$ and $V_k({\bf R})$ are the PESs of the $j$-th and $k$-th states, respectively,
and $\left\langle \exp \left[ 
\frac{i}{\hbar} U_{jk}({\bf R}_t ) \tau \right] \right\rangle$
corresponds to averaging 
$\exp \left[ 
\frac{i}{\hbar} U_{jk}({\bf R}_t ) \tau \right]$
over classical trajectories of the nuclear DOF, $\{ {\bf R}_t, {\bf P}_t \}$, whose initial state, $\{ {\bf R}_0, {\bf P}_0 \}$, is sampled from $\rho_n^{Cl} (0,{\bf R}_0, {\bf P}_0)$ [the classical limit of $\hat{\rho}_n (0)$] and whose dynamics is dictated by $V_j ({\bf R})$ 
(here, $\{ {\bf R}, {\bf P} \}$ correspond to the nuclear coordinates and momenta which are treated as classical variables).
IMT is obtained by assuming that the nuclear dynamics also satisfies Gaussian statistics. This translates into substituting $\left\langle \exp \left[ 
\frac{i}{\hbar} U_{jk}({\bf R}_t ) \tau \right] \right\rangle$ with its second-order cumulant expansion.\cite{mukamel_book,kubo85} Since the time integral is over a Gaussian integrand in this case, $k^{m}_{j\rightarrow k}(t)$ can be given in terms of an error function (erf):
\begin{align}
k^{m,IMT}_{j\rightarrow k}(t) =& \Gamma^2 \frac{\sqrt{2\pi}}{\sigma_{jk,t}} \Re \left( \erf \left[ \frac{\sigma_{jk,t}^2 \, t - i \hbar  \left\langle U_{jk} \right\rangle_t}{\sqrt{2} \, \hbar \sigma_{jk,t}} \right] \right)
   \label{eq:IMT}
\end{align}
where, $\langle U_{jk} \rangle_t = \langle U_{jk} ({\bf R}_t) \rangle$ and 
$\sigma_{jk,t}^2 = \langle U_{jk}^2 ({\bf R}_t) \rangle - \langle U_{jk} ({\bf R}_t) \rangle^2$.

 
The IMT expression in Eq. \ref{eq:IMT} can be further simplified by setting the upper limit of the time integral in Eq. \ref{eq:LSC_st_ht} to $\infty$. This is justified for times longer than the (typically rather short) lifetime of $\left\langle \exp \left[ -\frac{i}{\hbar} U_{jk}({\bf R}_t ) \tau \right] \right\rangle$, and leads for the following expression for the IMT NE-FGR rate coefficient, which was previously reported and discussed in Refs. \citenum{tong20,brian21}:  
\begin{align}
k^{m,LT-IMT}_{j\rightarrow k}(t) = \frac{\Gamma^2 }{\hbar} \sqrt{\frac{ 2\pi}{\sigma_{jk,t}^2}}     \exp \left[ -\frac{\left\langle U_{jk} \right\rangle_t^2}{2 \sigma_{jk,t}^2} \right] 
   \label{eq:LT-IMT}
\end{align}
In what follows, we will refer to the IMT expression in Eq. \ref{eq:LT-IMT} by LT-IMT, to denote that it is the long-time (LT) limit of the IMT rate coefficient in Eq. \ref{eq:IMT}.  
We also note that obtaining the cavity-modified IMT NE-FGR rate coefficients based on Eq \ref{eq:kshifted}, also requires calculating  
the IMT and LT-IMT approximations for $k^{m}_{j\rightarrow k}(t,\pm\omega_p)$, 
which can be obtained by replacing $\left\langle U_{jk} \right\rangle_t$ by 
$\left\langle U_{jk} \right\rangle_t \pm \hbar \omega_p$ in Eqs. \ref{eq:IMT} and \ref{eq:LT-IMT}.



\section{Cavity-modified rates for photo-induced charge transfer in the Carotenoid-Porphyrin-C$_{60}$ Triad dissolved in liquid Tetrahydrofuran}
\label{sec:cpc60}

\begin{figure}
    \centering
    \includegraphics[width=1\linewidth]{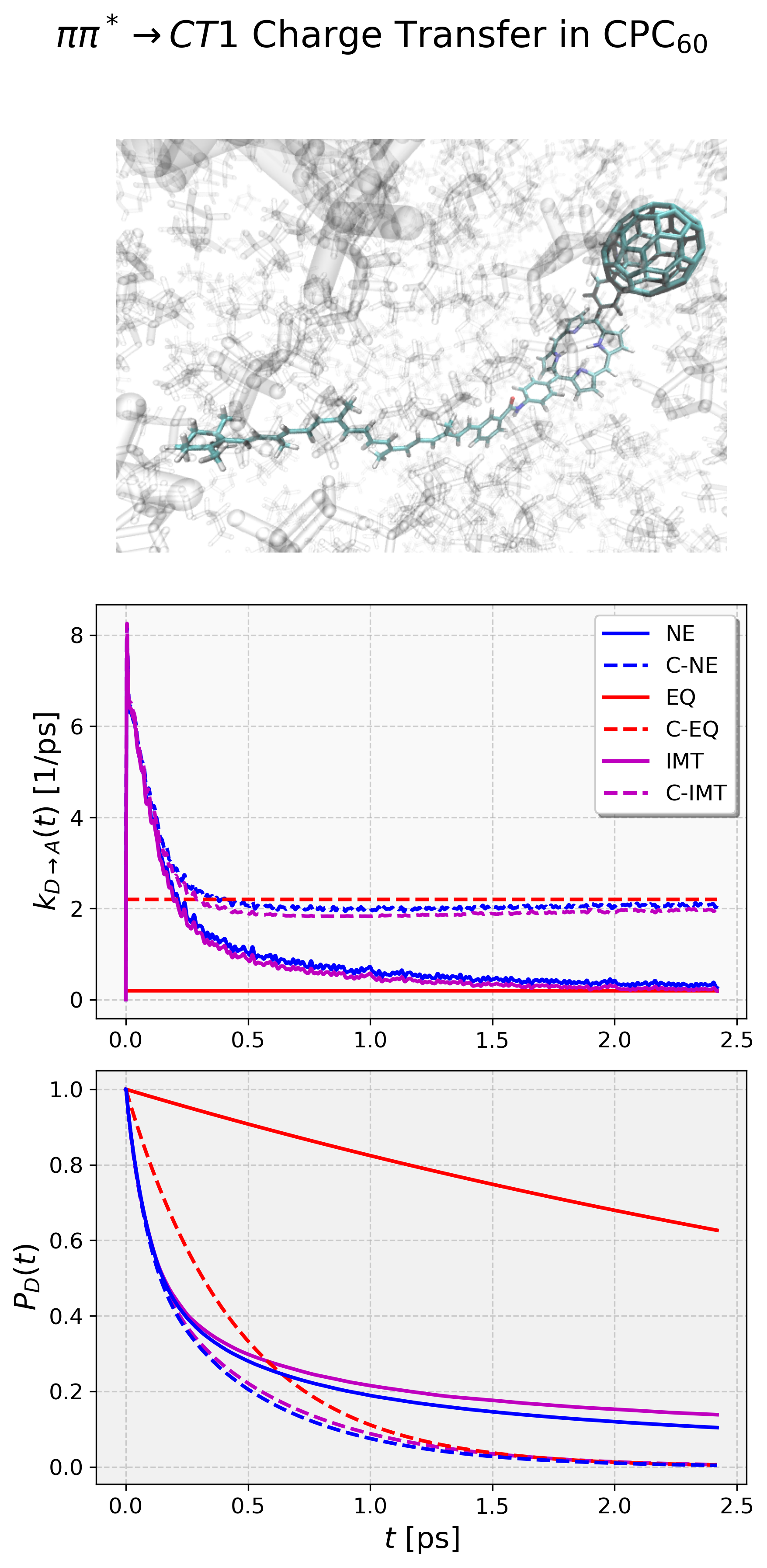}
    \caption{Top panel: A schematic view of CPC$_{60}$/THF.
    Middle panel: The NE-FGR rate coefficent, $k_{D\rightarrow A}(t)$, for the $\pi\pi^* \rightarrow$CT1 transition when CPC$_{60}$ is in its bent conformation. 
    Bottom panel: The population of the $\pi\pi^*$ donor state as a function of time, $P_D(t)$, simulated based on the 
    p-tcl-OD-QME, Eq. \ref{eq:p-tcl-OD-QME}.
    Shown are results based NE-FGR (NE), E-FGR (EQ) and the IMT approximation for NE-FGR (IMT). C indicates "in-cavity", while lack of C indicates "cavity-free". E.g., NE and C-NE corresponds to cavity-free and cavity-modified NE-FGR, respectively.}
    \label{fig:system_triad}
\end{figure}

The next two sections provide demonstrative applications of the theoretical framework developed in Sec. \ref{sec:theory}  to model systems.
We start out in this section, by demonstrating the effect of coupling to a cavity on photo-induced CT rates in the CPC$_{60}$
molecular triad dissolved in liquid THF 
(CPC$_{60}$/THF, see Fig. \ref{fig:system_triad}).\cite{sun18,hu20,brian21} We focus on the $\pi\pi^* \rightarrow$CT1 transition in this system, where $\pi\pi^*$ is the bright non-CT Porphyrin-localized excited state (CP$^*$C$_{60}$) and CT1 is the dark excited CT state (CP$^+$C$_{60}^-$). To this end, we employ the three-state harmonic model that was developed for this system in Ref. \citenum{brian21}. The triad is assumed to be in a bent conformation and to be impulsively photoexcited from the ground state to the $\pi\pi^*$ state, so that the initial nuclear configurations correspond to thermal equilibrium on the ground state PES. This choice is motivated by the fact that previous studies have found that the cavity-free $\pi\pi^* \rightarrow$CT1 NE-FGR rate is significantly enhanced relative to that predicted by E-FGR.\cite{hu20,brian21} This enhancement was traced back to the nonequilibrium  initial state of the nuclear DOF due to implusive photoexcitation from the ground state.\cite{hu20,brian21} In what follows, we investigate the effect of coupling to the cavity on the $\pi\pi^* \rightarrow$CT1 NE-FGR transition rate in CPC$_{60}$/THF. 

The results reported below were based on the following harmonic nuclear model Hamiltonians for the ground (CPC$_{60}$), donor (CP$^*$C$_{60}$) and acceptor (CP$^+$C$_{60}^-$):
\begin{equation}
\begin{split}
\hat H^n_D &= 
\hbar \omega_{DA} + 
\sum_{j=1}^N \left[ \frac{1}{2} \hat P_j^2 + \frac{1}{2} \omega_j^2 \hat R_j^2 \right] 
, \\
\hat H^n_A &= \sum_{j=1}^N \left[ \frac{1}{2} \hat P_j^2 + \frac{1}{2} \omega_j^2 (\hat R_j - R^{\text{eq}}_j)^2 \right] \\
\hat H^n_G &= E_G + \sum_{j=1}^N 
\left[ \frac{1}{2} \hat P_j^2 
+ \frac{1}{2} \omega_j^2 (\hat R_j + S_j)^2
\right]
\label{eq:triad_Hamiltonian}
\end{split}
\end{equation}

Here, $\{ \hat R_j, \hat P_j, \omega_j \,|\, j=1,\dots,N \}$ are the normal mode coordinates, momenta, and frequencies. $\{ R^{\text{eq}}_j \,|\, j=1,\dots,N \}$ and $\{ S_j \,|\, j=1,\dots,N \}$ are the donor-to-acceptor and donor-to-ground shifts in the equilibrium geometry along the normal mode coordinates. The parameters $E_G$ and $\hbar\omega_{DA}$ are shifts in equilibrium geometry energies  
of the ground state and donor state relative to the acceptor state, respectively. 

The model Hamiltonians in Eq. \ref{eq:triad_Hamiltonian} were parametrized in Ref. \citenum{brian21} based on inputs from all-atom molecular dynamics simulations. More specifically, the values of $\{ \omega_j, R_j^{eq}, S_j \}$ we use here correspond to the ones obtained via the procedure referred to in Ref. \citenum{brian21} as {\em model 2}. 
The coupling strength to the cavity mode was set to $\hbar g_p =10$meV, which is comparable to the cavity-free $\pi\pi^*$-CT1 electronic coupling coefficient, $V_{DA}^n = \Gamma = 24$meV, which was previously obtained from electronic structure calculations.\cite{sun18,hu20,brian21}  The cavity frequency was set to $\hbar \omega_p = 591$meV,  which is the value that maximizes the cavity-induced enhancement of E-FGR for this system based on the procedure proposed in Ref. \citenum{saller_cavity-modified_2023}. Finally, $\hbar \omega_{DA}$ was set to 1014 meV\cite{hu20,brian21} and the temperature was set to 300K.

Importantly, $C^n_{jk}(t,\tau)$ can be obtained in closed form for the model under consideration:\cite{lai_simulating_2021}

\begin{widetext}
\begin{align}
    &C^n_{DA}(t, \tau) = \Gamma^2 \exp\left(i \, \omega_{DA} \, \tau\right) \label{eq:nefgr_cf1} \\
    &\exp\left(\sum_j \left[ \frac{-i \, R^{\text{eq}}_j \, \omega_j}{\hbar \, \sinh\left(\frac{\beta \hbar \omega_j}{2}\right)} \sin\left(\frac{\tau \, \omega_j}{2}\right) \left(R^{\text{eq}}_j \, \sinh\left(\frac{\beta \hbar - i \, \tau}{2} \omega_j\right) + S_j \left(\sinh\left(\frac{\beta \hbar - i (t - \tau)}{2} \omega_j\right) + \sinh\left(\frac{\beta \hbar + i (t - \tau)}{2} \omega_j\right)\right)\right)\right]\right) \notag \\
    &C^n_{AD}(t, \tau) = \Gamma^2  \exp\left(-i \, \omega_{DA} \, \tau\right) \label{eq:nefgr_cf2} \\
    &\exp\left(\sum_j \left[ \frac{i \, R^{\text{eq}}_j \, \omega_j}{\hbar \, \sinh\left(\frac{\beta \hbar \omega_j}{2}\right)} \sin\left(\frac{\tau \, \omega_j}{2}\right) \left(-R^{\text{eq}}_j \, \sinh\left(\frac{\beta \hbar - i \, \tau}{2} \omega_j\right) + \left(R^{\text{eq}}_j + S_j\right) \left(\sinh\left(\frac{\beta \hbar - i (t - \tau)}{2} \omega_j\right) + \sinh\left(\frac{\beta \hbar + i (t - \tau)}{2} \omega_j\right)\right)\right)\right]\right) \notag
\end{align}    
\end{widetext}
The main quantities that determine the IMT rates can also be obtained in closed form:
\begin{align}
    \langle U_{DA} \rangle_t &= \hbar \omega_{DA} \label{eq:imt_avg1} \\
    &- \sum_j \left[\frac{1}{2} \omega_j^2 \left(R^{\text{eq}}_j\right)^2 + \omega_j^2 R^{\text{eq}}_j S_j \cos\left( \omega_j t\right)\right] \notag \\
    \langle U_{AD} \rangle_t &=-\hbar\omega_{DA} \label{eq:imt_avg2} \\
    &+\sum_j \left[\omega_j^2 R^{\text{eq}}_j \left(R^{\text{eq}}_j + S_j\right) \cos\left(\omega_j t\right) -\frac{1}{2} \omega_j^2 \left(R^{\text{eq}}_j\right)^2 \right] \notag
\end{align}
\begin{align}
     \sigma_{DA,t}^2=\sigma_{AD,t}^2 = \sum_{j=1}^N \frac{\omega_j^2 \left(R^{\text{eq}}_j\right)^2}{\beta} \label{eq:imt_var}
\end{align}

Finally, we note that closed-form expressions for the donor-to-acceptor E-FGR rate constant, $k_{D\rightarrow A}^{eq}$, can be found in Ref.  \citenum{saller_cavity-modified_2023}, and 
that $k_{A\rightarrow D}^{eq}$ can be obtained from $k_{D\rightarrow A}^{eq}$ by using detailed balance, $k_{A\rightarrow D}^{eq}=\exp(-\beta\hbar\omega_{DA} )k_{D\rightarrow A}^{eq}$.

The effect of coupling to the cavity on the $\pi\pi^* \rightarrow \text{CT}1$ CT dynamics in CPC$_{60}$/THF simulated based on the p-tcl-OD-QME, Eq. \ref{eq:p-tcl-OD-QME},
is shown in Fig.~\ref{fig:system_triad}. The lower panel of this figure depicts the population of the donor state as a function of time, $P_D (t)$, while the middle panel shows the CT rate coefficient as a function of time, $k_{D \rightarrow A} (t)$. 


A close inspection of Fig. \ref{fig:system_triad} gives rise to the following noteworthy observations:
\begin{itemize}
\item Both cavity-free and cavity-modified NE-FGR rate coefficients, $k_{D \rightarrow A} (t)$, reach a plateau following a nonequilibrium transient period, with the plateau value given by the corresponding E-FGR rate constant. This is the expected behavior in condensed-phase systems, where the nonequilibrium initial state equilibrates (often on a time scale comparable to that of CT). Since the effect of the cavity on the E-FGR rate constant was already reported and analyzed in Ref. \citenum{saller_cavity-modified_2023}, we will focus here on the effect of the cavity on the nonequilibrium portion of the dynamics. 
\item The cavity-free NE-FGR rate of the $\pi\pi^* \rightarrow \text{CT}1$ transition in CPC$_{60}$/THF is seen to be much faster than the rate predicted by E-FGR. 
This behavior has been reported previously, and traced back to the large deviation between the nuclear configurations in the ground state and the $\pi\pi^*$ state, and the fact that the ground state nuclear configurations are in closer proximity to the crossing region between the $\pi\pi^*$ and CT1 PESs, which leads to the observed enhancement in the CT rate.\cite{hu20}
However, placing the CPC$_{60}$/THF system in a resonant cavity makes the difference between NE-FGR and E-FGR rates smaller. This can be traced back to the fact that the cavity-modified enhancement of the NE-FGR rates involves $k^m_{\pi\pi^*\rightarrow {CT1}}(t, \pm \omega_p)$, that corresponds to shifting the $\pi\pi^*$ PES by $\pm \hbar \omega_p$, which in turn 
moves the crossing point between the $\pi\pi^*$ and CT1 PESs relative to the ground state nuclear configurations.
As a result, the ground state equilibrium nuclear configurations 
end up being further away from the crossing point between the {\em dressed} $\pi\pi^*$ and CT1 PESs, 
thereby decreasing the nonequilibrium effects, as measured by the difference between E-FGR and NE-FGR rates.  
\item The IMT approximation is observed to accurately reproduce the NE-FGR CT dynamics. Thus, the high temperature and short time limits it is based on appear to be justified for this prototypical photo-induced CT reaction at room temperature.      
\end{itemize}

\section{Cavity-modified rates for the Garg-Onuchic-Ambegaokar (GOA) model for charge transfer in the condensed phase}
\label{sec:model}

In this section, we apply the theoretical framework developed in Sec. \ref{sec:theory} to the 
GOA model for CT in the condensed phase.\cite{garg85} To this end, we 
compare and contrast the CT dynamics inside and outside of the cavity based on the p-tcl-OD-QME, Eq. \ref{eq:p-tcl-OD-QME}. We do so in both the normal and inverted regions, for different nonequilibrium initial states and at different temperature and frictions (see below).  



  The cavity-free Hamiltonian for the GOA model is given by:
\begin{eqnarray}
			\hat{H}^m &=& \Gamma \hat{\sigma}_x +\frac{\hbar\omega_{DA}}{2}\hat{\sigma}_z + \frac{\hat{P}^2_y}{2} + \frac{1}{2}\Omega^2(\hat{y}+y_0\sigma_z)^2
\nonumber \\
			&&+ \sum^{N-1}_{\alpha=1}\left[\frac{\hat{p}_\alpha^2}{2}+\frac{1}{2}\omega_\alpha^2\left(\hat{x}_\alpha+\frac{c_\alpha\hat{y}}{\omega_\alpha^2}\right)^2\right]
            \label{eq_GOA_H_l}
\end{eqnarray}
Here, $\hat{\sigma}_x = | D \rangle \langle A| + | A \rangle \langle D|$ and $\hat{\sigma}_z = | D \rangle \langle D| - | A \rangle \langle A|$, where $| D \rangle$ and $ | A \rangle$ are the donor and acceptor electronic states; $V_{DA}^n = \Gamma$ is the electronic coupling coefficient (constant within the Condon approximation); $- \hbar \omega_{DA}$ is the donor-to-acceptor CT reaction energy; $\hat{y}$, $\hat{P}_y$, and $\Omega$ are the mass-weighted primary mode coordinate, momentum, and frequency, respectively; $\{ \hat{x}_j, \hat{p}_j, \omega_j | j = 1,...,N-1\}$ are the mass-weighted coordinates, momenta and frequencies, respectively, of the secondary bath modes; $\{ c_j | j = 1,...,N-1\}$ are the coupling coefficients between the secondary modes and the primary mode; $2 y_0$ is the shift in equilibrium geometry between the donor and acceptor states along the primary mode, such that the reorganization energy is given by $E_r = 2 \Omega^2 y_0^2$. 
%
Finally, the bath of secondary modes is described by an Ohmic spectral density with exponential cutoff:
\begin{equation}
    J(\omega) = \frac{\pi}{2} \sum_{j=1}^{N-1} \frac{c_j^2}{\omega_j} \delta (\omega - \omega_j)
    = \eta \omega e^{-\omega/\omega_c}~~.
    \label{eq:SD}
\end{equation}
Here, $\eta$ is the friction coefficient, which determines the coupling strength between the primary and secondary modes, and $\omega_c$ is the cutoff frequency (both of which adjustable parameters).
The form of the Hamiltonian in Eq. \ref{eq_GOA_H_l} requires representing the continuous spectral density in Eq. (\ref{eq:SD}) in terms of a set of discrete secondary bath modes. 
To this end, we follow the procedure outlined in Refs. \citenum{makri99c,chowdhury21}, with 
the number of secondary modes set to $N-1 = 200$

The nonequilibrium initial state of the nuclear DOF (see Eq. \ref{eq:initstate}) is obtained by assuming that the ground state equilibrium geometry is shifted by $s$ along the primary mode coordinate relative to the equilibrium geometry of the D state, such that 
\begin{equation}
\hat{\rho}_n (0) = \frac{e^{-\beta\hat{H}^n_G}}{\Tr_n\left[e^{-\beta\hat{H}^n_G}\right]}~~,
\end{equation}
where 
\begin{eqnarray}
\hat{H}^n_G &=& E_G + \frac{\hat{P}^2_y}{2} + \frac{1}{2}\Omega^2(\hat{y}+y_0 + s)^2
\nonumber \\
&&+ \sum^{N-1}_{\alpha=1}\left[\frac{\hat{p}_\alpha^2}{2}+\frac{1}{2}\omega_\alpha^2\left(\hat{x}_\alpha+\frac{c_\alpha\hat{y}}{\omega_\alpha^2}\right)^2\right]~~.
\label{eq_GOA_G} 
\end{eqnarray}
Here, $E_G < 0$ is the energy of the ground state at its equilibrium geometry relative to the acceptor energy in its equilibrium geometry.

The fact that the primary and secondary modes are bilineraly coupled implies that the nuclear coordinates can be cast in terms of the corresponding normal modes instead of in terms of the primary mode and secondary modes.\cite{sun16,sun16a} Doing so allows for casting the GOA Hamiltonian in the following alternative and yet completely equivalent form:
\begin{equation}	\hat{H}^m=\hat{H}^{n}_D\dyad{D}+\hat{H}^{n}_A\dyad{A}+ \Gamma [\dyad{D}{A}+\dyad{A}{D}]~~.
\label{eq:overall_HGOA}
\end{equation}
where, 
\begin{equation}
\begin{split}
\hat H^n_D &= 
\hbar \omega_{DA} + 
\sum_{j=1}^N \left[ \frac{1}{2} \tilde P_j^2 + \frac{1}{2} \tilde \omega_j^2 \tilde R_j^2 \right] 
, \\
\hat H^n_A &= \sum_{j=1}^N \left[ \frac{1}{2} \tilde P_j^2 + \frac{1}{2} \tilde \omega_j^2 (\tilde R_j - \tilde R^{eq}_j)^2 \right]
\label{eq:GOA_Hamiltonian}
\end{split}
\end{equation}
Here, $\{ \tilde R_j, \tilde P_j, \tilde \omega_j | j=1,...,N \}$ are the normal mode coordinates, momenta, and frequencies, and $\{ \tilde R^{eq}_j| n=1,...,N \}$ are the donor-to-acceptor shifts in the equilibrium geometry along the normal mode coordinates.
%
%
%
The cavity-free nuclear Hamiltonian for the ground electronic state can also be cast in terms of the normal modes coordinates and momenta:
\begin{align}
\hat{H}^n_G=E_G+ \sum_k 
\left[ \frac{1}{2}\tilde{P}^2_k 
+ \frac{1}{2} \tilde \omega_j^2(\tilde {R}_j+\tilde s_j)^2
\right]
\label{eq:hq1}
\end{align}
where
$\{ \tilde s_j| n=1,...,N \}$ are the donor-to-ground 
shifts in the equilibrium geometry along the normal mode coordinates. The closed form expressions for the NE-FGR correlation functions, IMT average energy gaps, and IMT variances derived for CPC$_{60}$ 
can be used for the GOA model by replacing $\{ \omega_j, ~ S_j, ~ R^{eq}_j\}$ with $\{ \tilde \omega_j, ~\tilde s_j, ~\tilde R^{eq}_j\}$ (see Eqs. \ref{eq:nefgr_cf1}-\ref{eq:imt_var}). Then the rates and populations can be calculated as explained in Sec. \ref{sec:theory}





\begin{figure}
    \centering
    \includegraphics[width=1\linewidth]{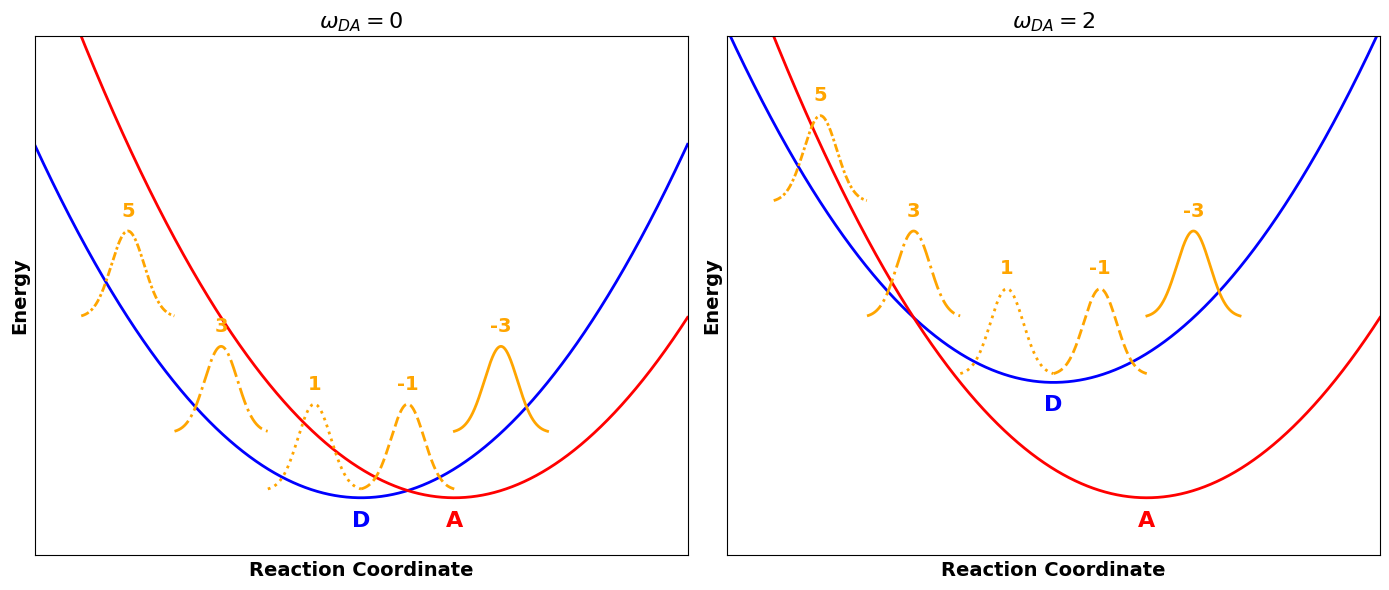}
    \caption{A schematic view of the nonequilibrium initial states under consideration for the GOA model, in the normal (left panel) and inverted (right panel) regions. The donor (blue) and acceptor (red) PESs are plotted along the GOA model’s primary mode coordinate. The numbers (-3, -1, 1, 3, 5) correspond to the different values of the shift, $s$, relative to the donor's equilibrium geometry (orange, see Eq. \ref{eq_GOA_G}).
    }
    \label{fig:ct_schematic}
\end{figure}

\begin{figure}
    \centering
    \includegraphics[width=1\linewidth]{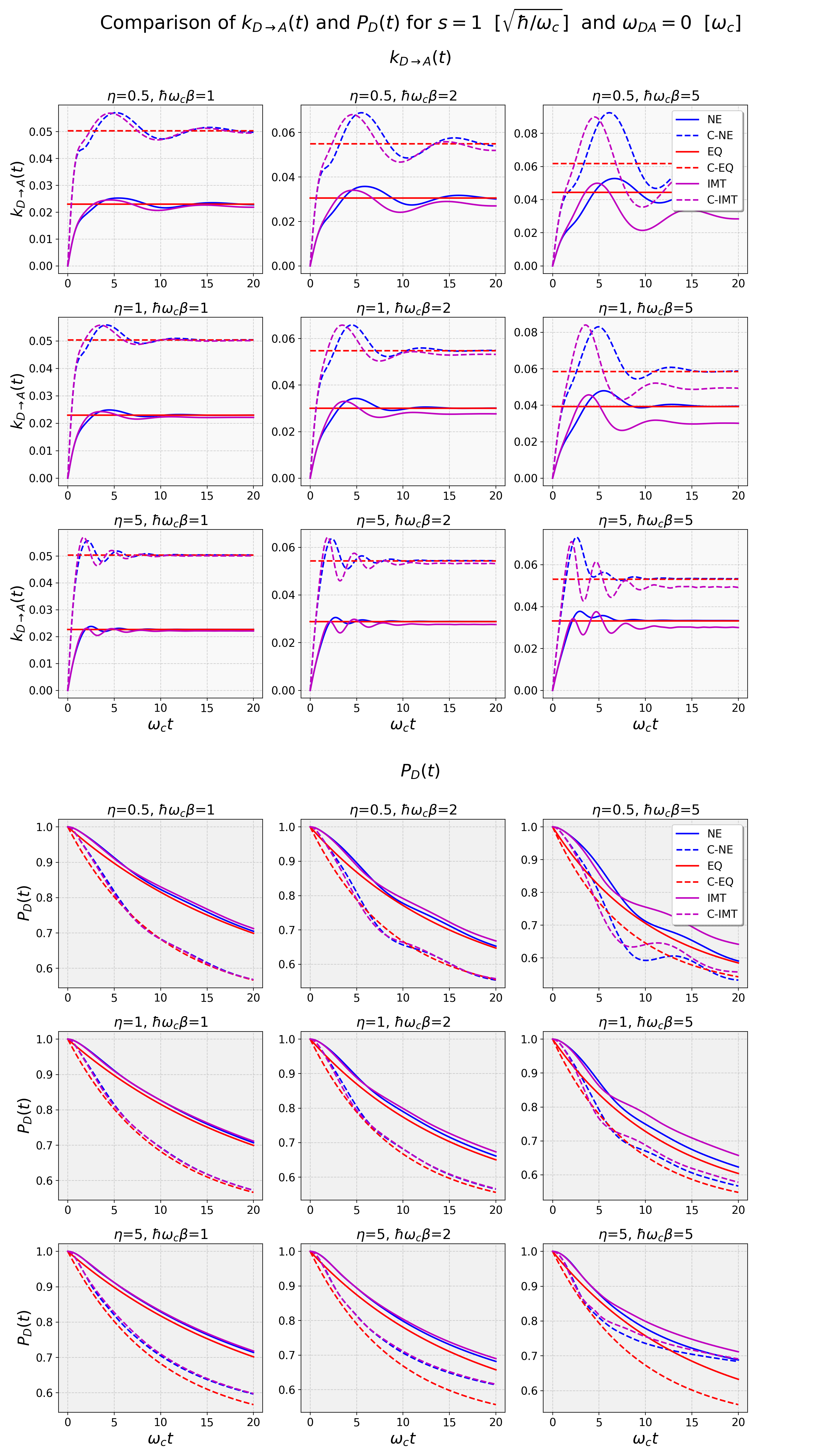}
    \caption{Comparison of donor-to-acceptor transition rates, $k_{D\rightarrow A}(t)$, (top panels) and donor population, $P_D ( t )$, (bottom panels) as a function of time, at different temperatures and frictions, for the GOA model with $s/\sqrt{\hbar / \omega_c}=1.0$ and $\omega_{DA}/\omega_c=0.0$. Shown are results based NE-FGR (NE), E-FGR (EQ) and the IMT approximation for NE-FGR (IMT). C indicates "in-cavity", while lack of C indicates "cavity-free". For example, NE and C-NE corresponds to cavity-free and cavity-modified NE-FGR, respectively.}
    \label{fig:ks1wda0}
\end{figure}

\begin{figure}
    \centering
    \includegraphics[width=1\linewidth]{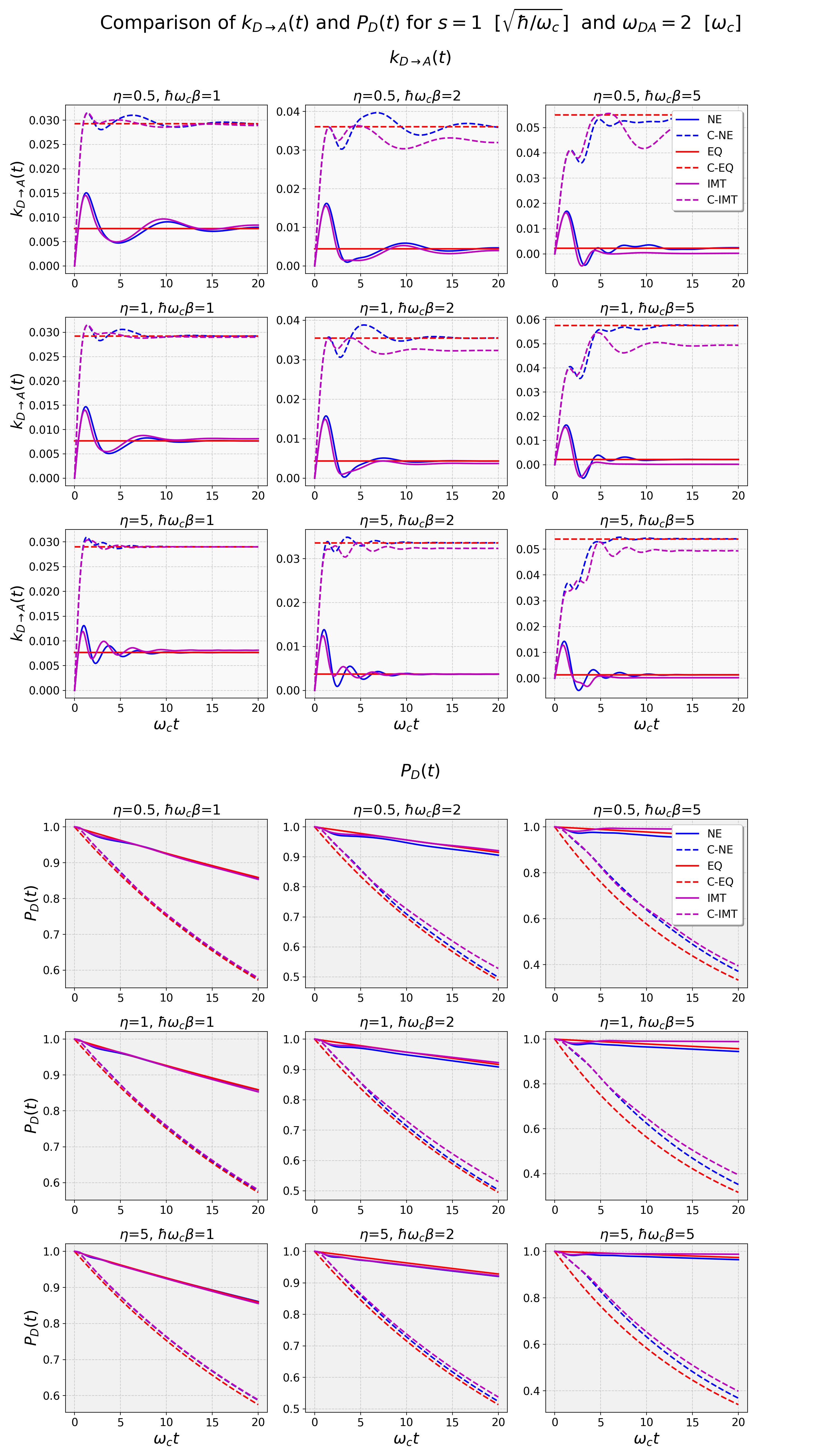}
    \caption{Same as Fig. \ref{fig:ks1wda0} for the GOA model with $s/\sqrt{\hbar / \omega_c}=1.0$ and $\omega_{DA}/\omega_c=2.0$. 
    }
    \label{fig:ks1wda2}
\end{figure}

\begin{figure}
    \centering
    \includegraphics[width=1\linewidth]{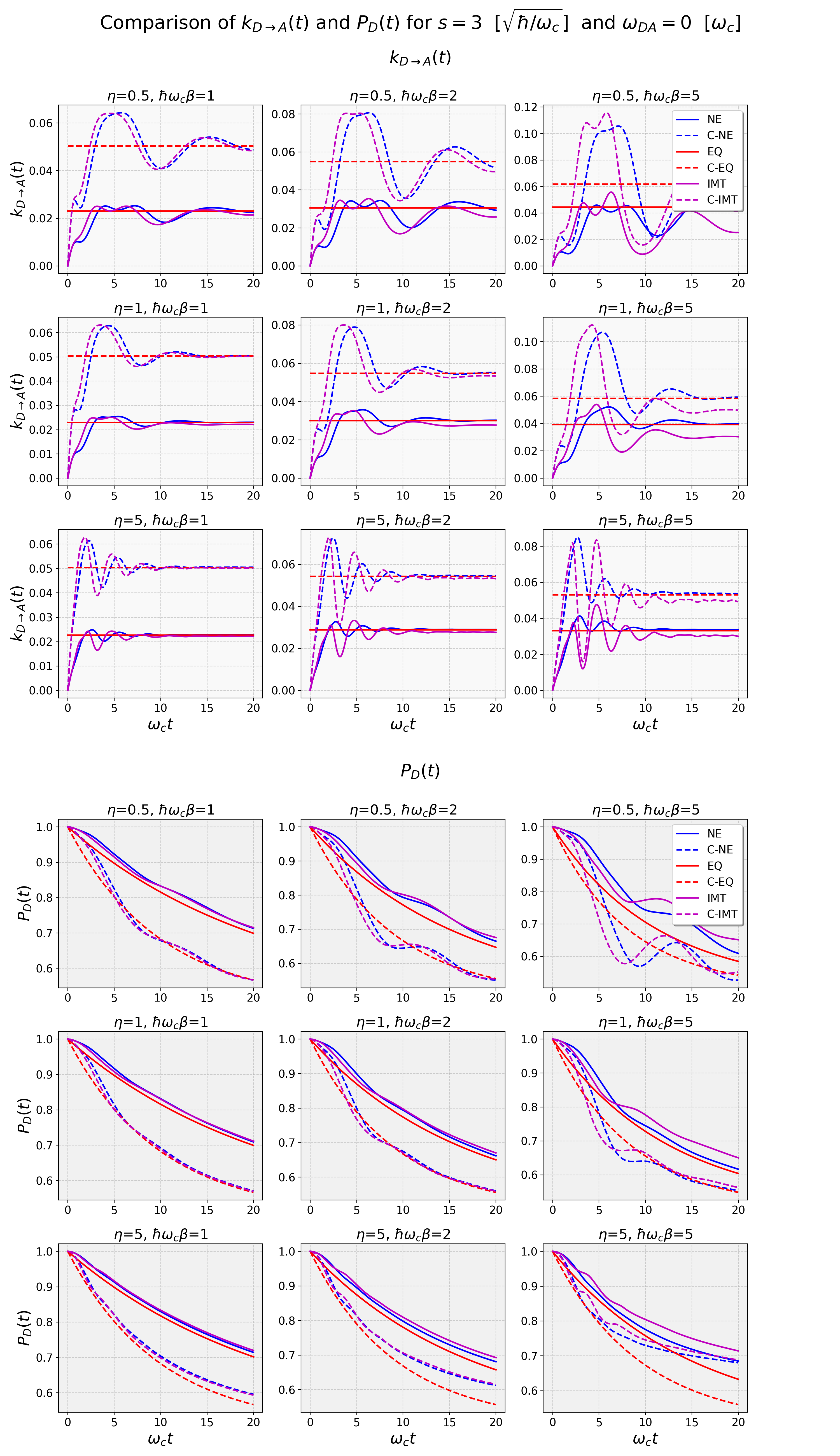}
    \caption{Same as Fig. \ref{fig:ks1wda0} for the GOA model with $s/\sqrt{\hbar / \omega_c}=3.0$ and $\omega_{DA}/\omega_c=0.0$. 
    }
    \label{fig:ks3wda0}
\end{figure}

\begin{figure}
    \centering
    \includegraphics[width=1\linewidth]{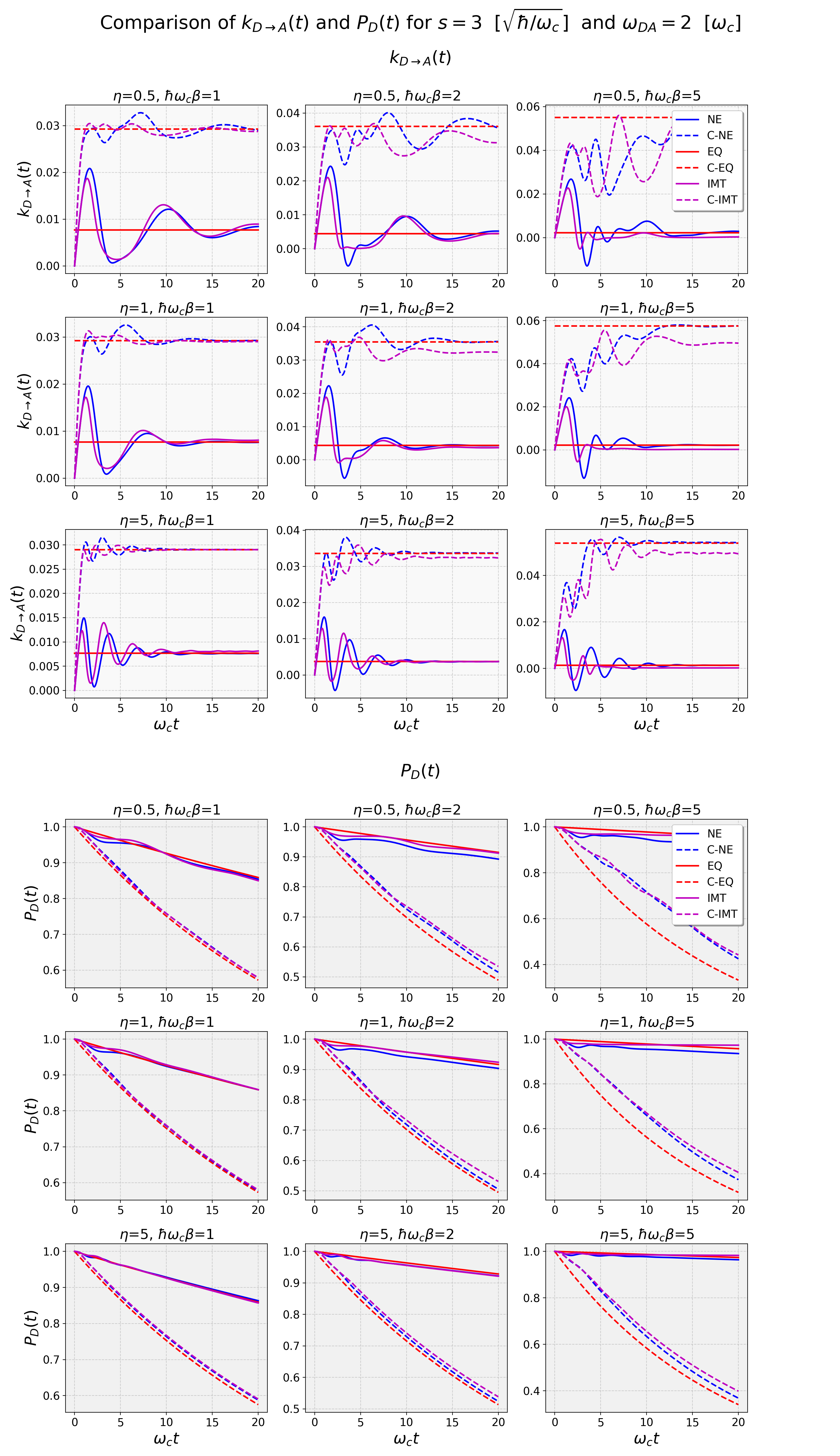}
    \caption{Same as Fig. \ref{fig:ks1wda0} for the GOA model with $s/\sqrt{\hbar / \omega_c}=3.0$ and $\omega_{DA}/\omega_c=2.0$. 
    }
    \label{fig:ks3wda2}
\end{figure}

\begin{figure}
    \centering
    \includegraphics[width=1\linewidth]{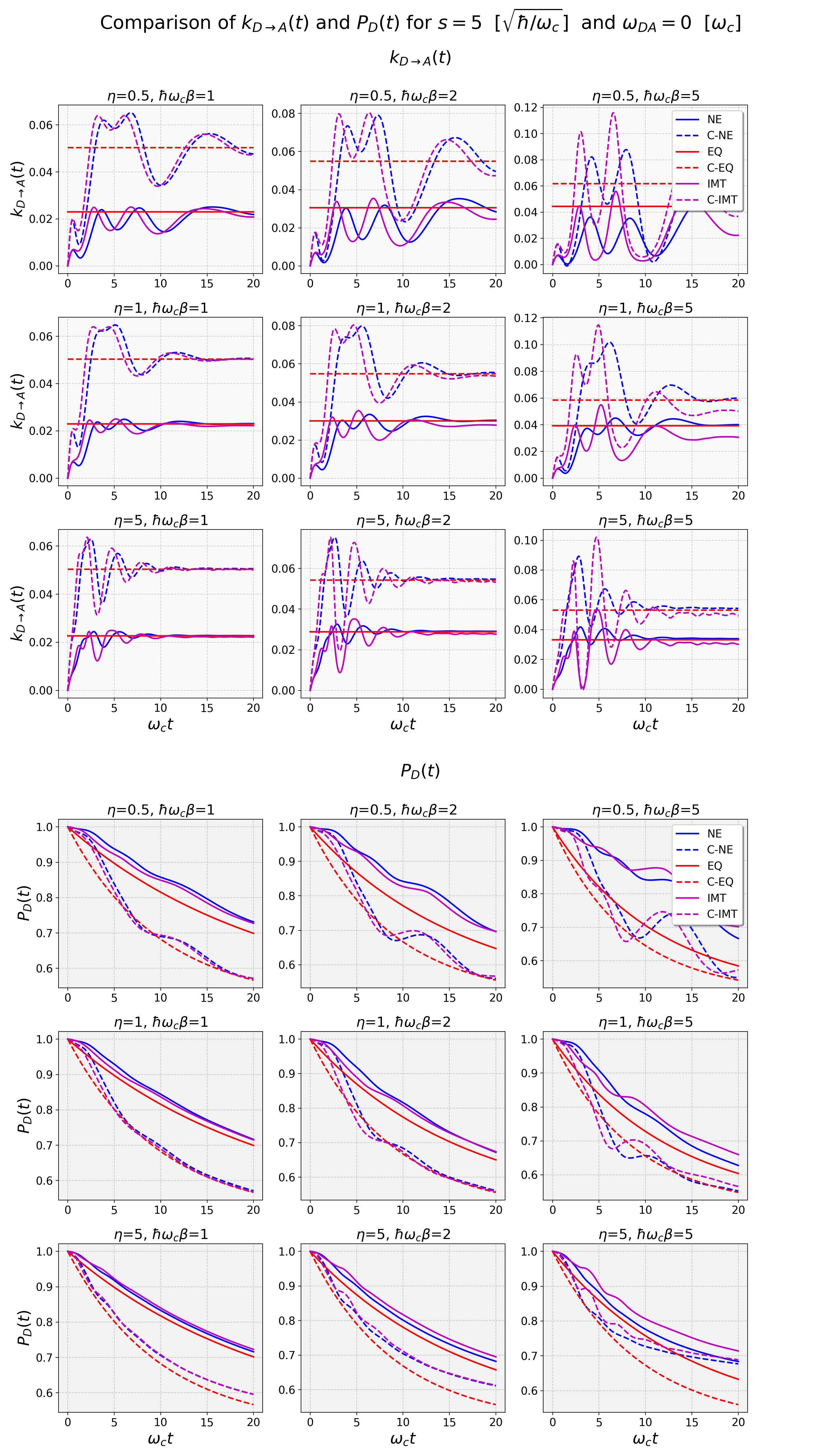}
    \caption{Same as Fig. \ref{fig:ks1wda0} for the GOA model with $s/\sqrt{\hbar / \omega_c}=5.0$ and $\omega_{DA}/\omega_c=0.0$. 
    }
    \label{fig:ks5wda0}
\end{figure}

\begin{figure}
    \centering
    \includegraphics[width=1\linewidth]{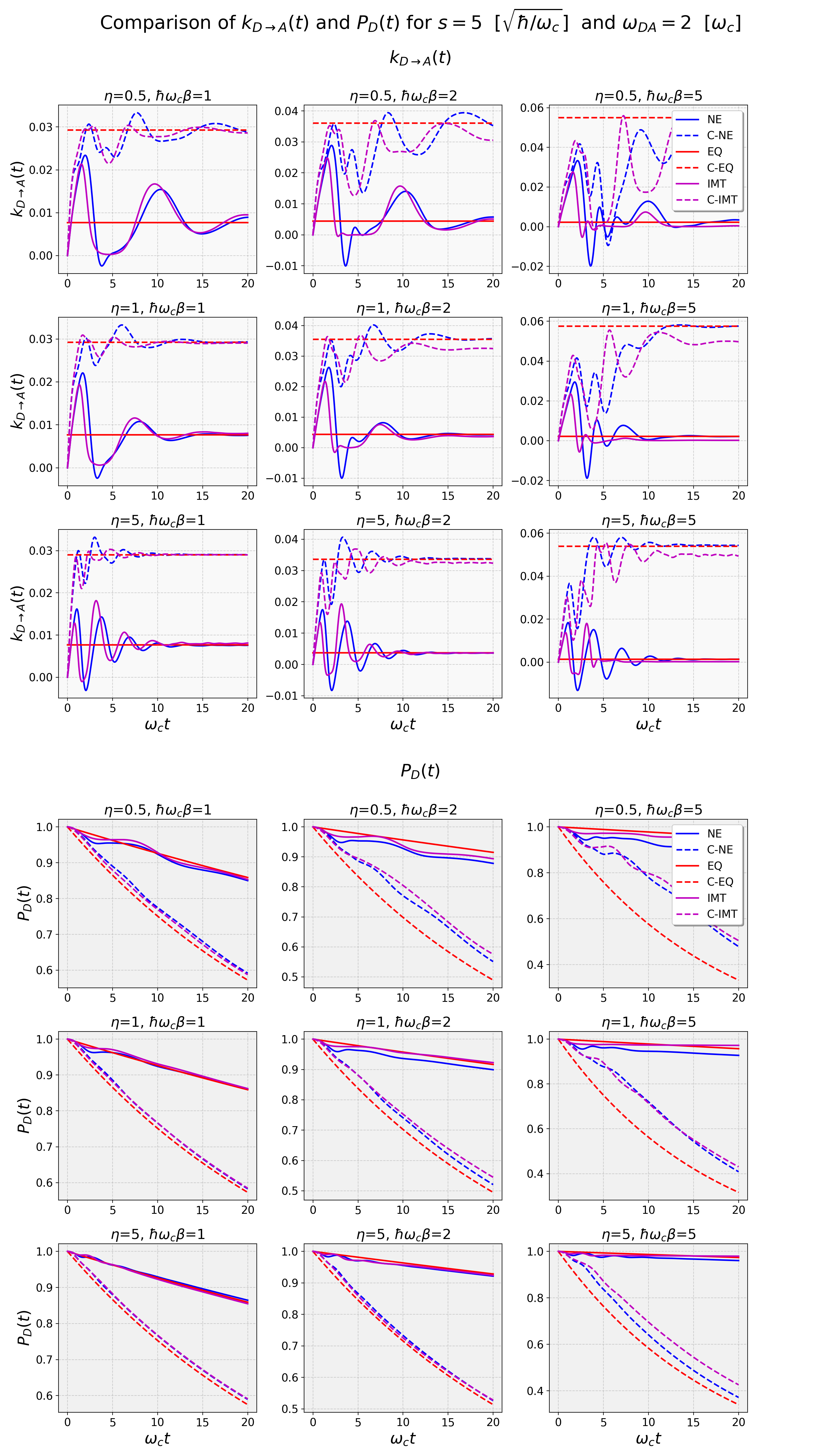}
    \caption{Same as Fig. \ref{fig:ks1wda0} for the GOA model with $s/\sqrt{\hbar / \omega_c}=5.0$ and $\omega_{DA}/\omega_c=2.0$. 
    }
    \label{fig:ks5wda2}
\end{figure}

\begin{figure}
    \centering
    \includegraphics[width=1\linewidth]{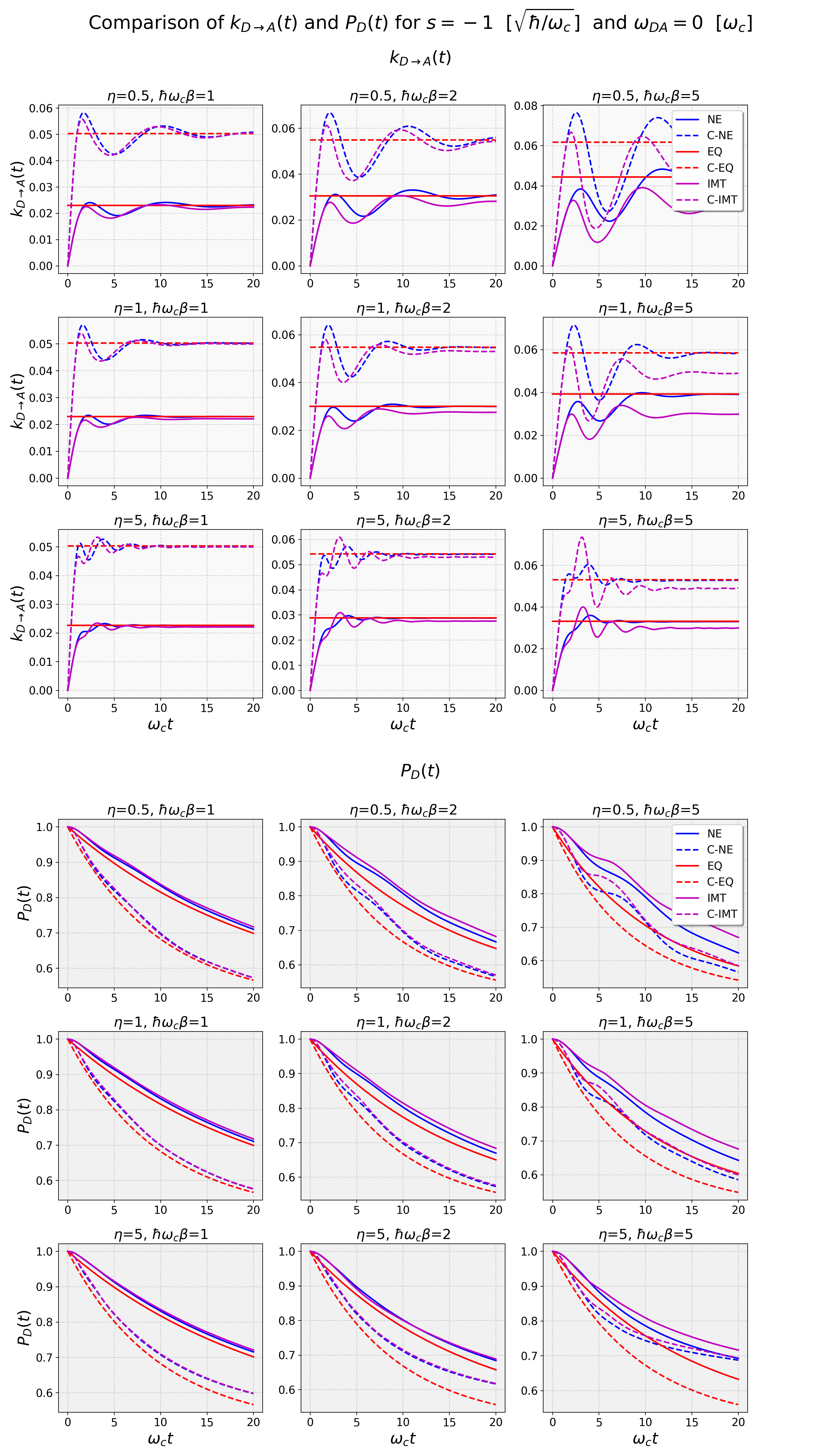}
    \caption{Same as Fig. \ref{fig:ks1wda0} for the GOA model with $s/\sqrt{\hbar / \omega_c}=-1.0$ and $\omega_{DA}/\omega_c=0.0$. 
    }
    \label{fig:ks-1wda0}
\end{figure}

\begin{figure}
    \centering
    \includegraphics[width=1\linewidth]{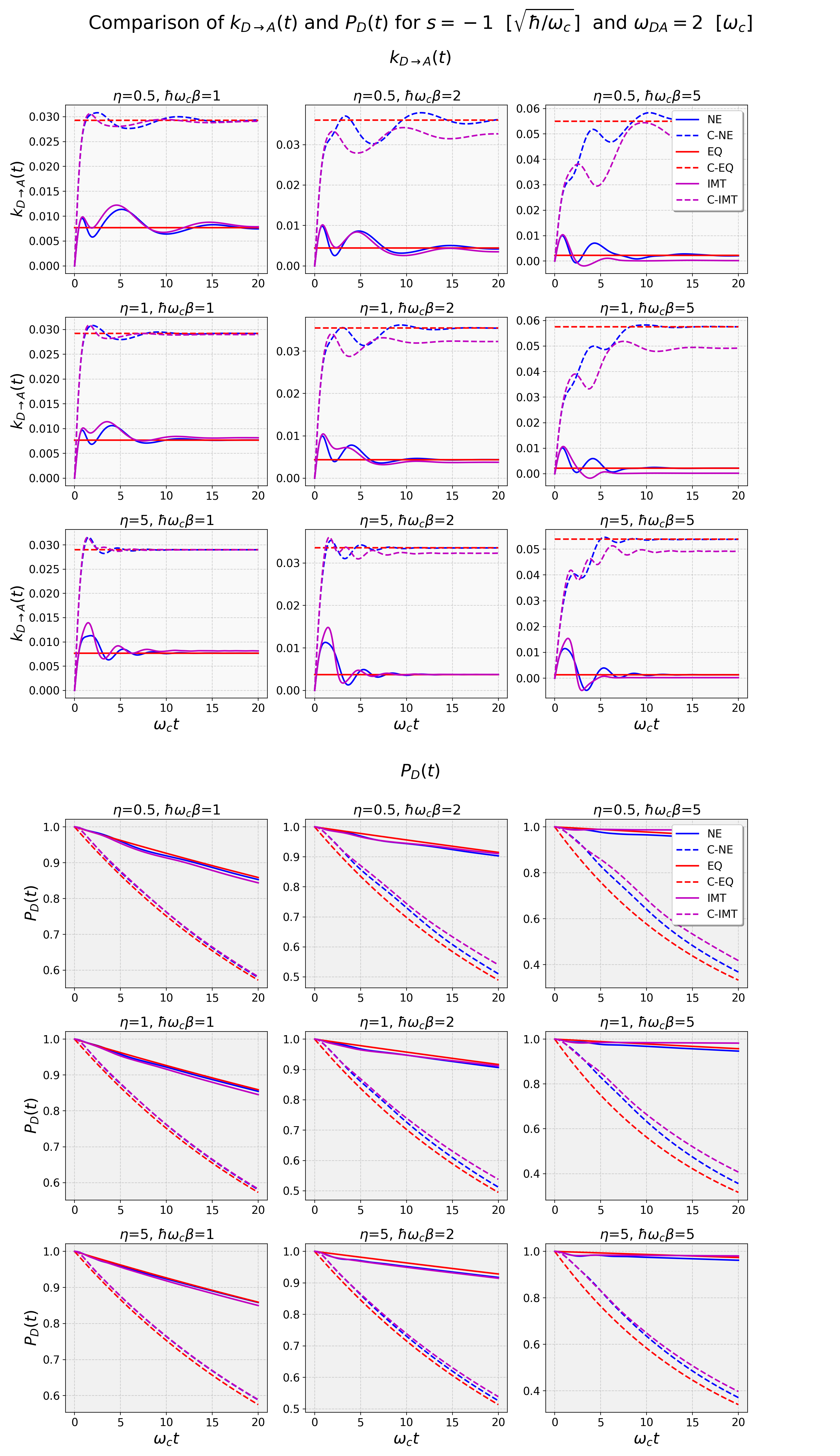}
    \caption{Same as Fig. \ref{fig:ks1wda0} for the GOA model with $s/\sqrt{\hbar / \omega_c}=-1.0$ and $\omega_{DA}/\omega_c=2.0$. 
    }
    \label{fig:ks-1wda2}
\end{figure}

\begin{figure}
    \centering
    \includegraphics[width=1\linewidth]{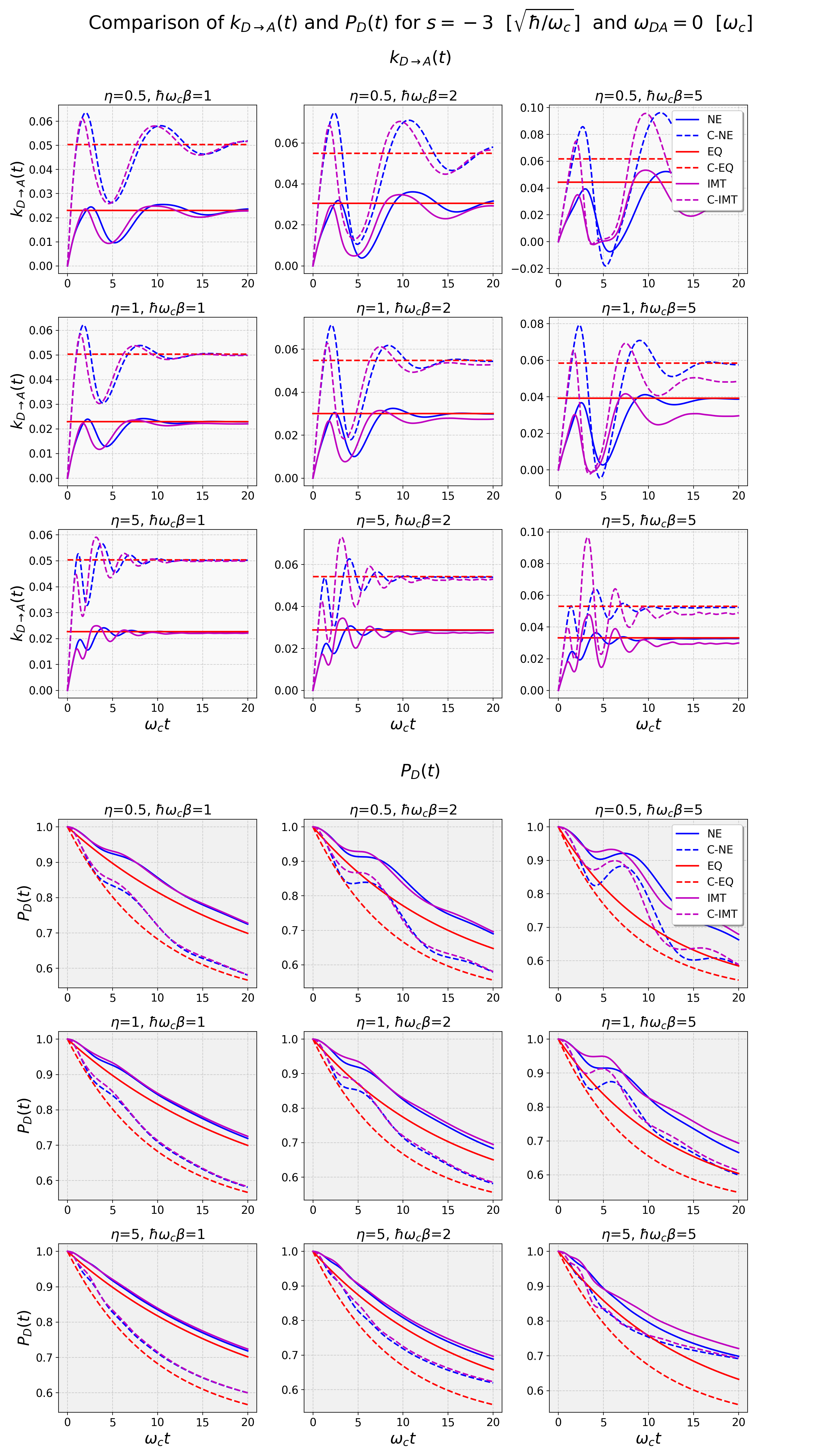}
    \caption{Same as Fig. \ref{fig:ks1wda0} for the GOA model with $s/\sqrt{\hbar / \omega_c}=-3.0$ and $\omega_{DA}/\omega_c=0.0$. 
}
    \label{fig:ks-3wda0}
\end{figure}

\begin{figure}
    \centering
    \includegraphics[width=1\linewidth]{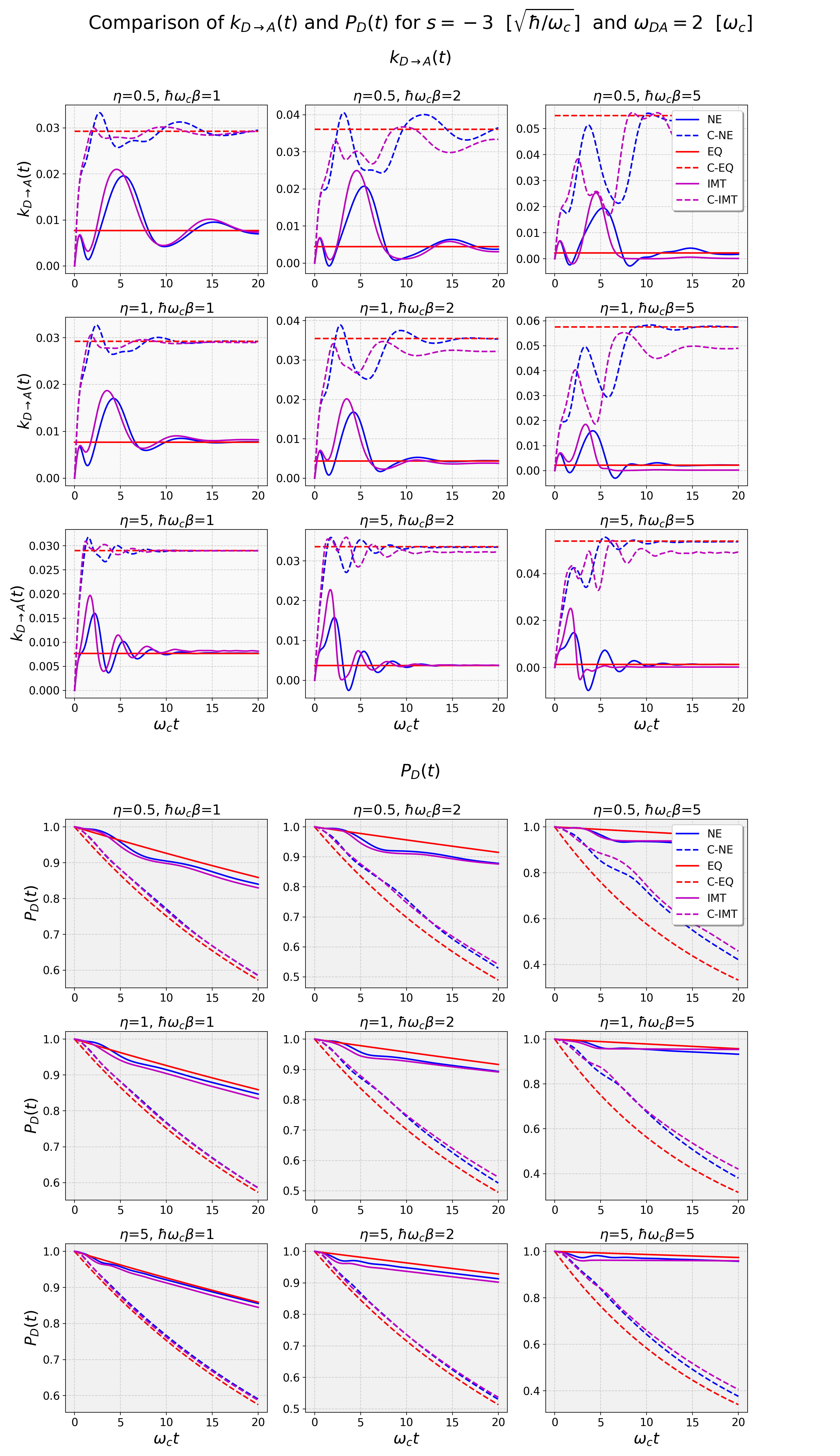}
    \caption{Same as Fig. \ref{fig:ks1wda0} for the GOA model with $s/\sqrt{\hbar / \omega_c}=-3.0$ and $\omega_{DA}/\omega_c=2.0$. 
    }
    \label{fig:ks-3wda2}
\end{figure}

The results reported below were obtained for the parameters in Table \ref{tab:parameters}.
The two values of $\omega_{DA}$ correspond to the normal region ($\omega_{DA}/\omega_c=0.0$) and inverted region ($\omega_{DA}/\omega_c=2.0$). 
For each case, we consider the CT dynamics inside and outside the cavity for several different initial nonequilibrium states corresponding to different values of $s$ ($s/\sqrt{\hbar / \omega_c}=\{-3.0, -1.0, 1.0, 3.0, 5.0 \}$, see Fig. \ref{fig:ct_schematic}). The dependence of the cavity-free and cavity-modified CT dynamics on $\eta$ and $\beta$ ($\beta \hbar \omega_c =\{5.0,2.0,1.0\}$ and $\eta = \{0.5,1.0,5.0 \}$)
for each initial state is shown in Figs. \ref{fig:ks1wda0} \ref{fig:ks3wda0}, \ref{fig:ks5wda0}, \ref{fig:ks-1wda0}, \ref{fig:ks-3wda0} (normal region) and \ref{fig:ks1wda2}, \ref{fig:ks3wda2}, \ref{fig:ks5wda2}, \ref{fig:ks-1wda2},\ref{fig:ks-3wda2} (inverted region). 

Similarly to the CPC$_{60}$/THF model system in the previous section, the value of the coupling strength to the cavity mode,  $\hbar g_p$, picked is the same as the cavity-free donor-acceptor coupling, $V_{DA}^n = \Gamma$ (see Table \ref{tab:parameters}), and the cavity frequency, $\omega_p$, was determined so as to maximize the cavity-induced enhancement of E-FGR of the system based on the procedure proposed in Ref. \citenum{saller_cavity-modified_2023} (see Table \ref{tab:omega_p}). 




\begin{table}[h]
\centering
\begin{tabular}{|c|c|}
\hline
$g_p / \omega_c$ & 1.0 \\
$\Gamma / \hbar\omega_c$ & 1.0 \\
$\Omega / \omega_c$ & 0.5 \\
$\omega_{DA} / \omega_c$ & 0.0, 2.0 \\
$k_B T / \hbar \omega_c$ & 1.0, 0.5, 0.2 \\
$s / \sqrt{\hbar / \omega_c}$ & $-3.0$, $-1.0$, $1.0$, $3.0$, $5.0$ \\
$\eta / \omega_c$ & 0.5, 1.0, 5.0 \\
$y_0 / \sqrt{\hbar / \omega_c}$ & 1.0 \\
$N_s$ & 200 \\
\hline
\end{tabular}
\caption{Simulation parameters.}
\label{tab:parameters}

\vspace{1em}

\begin{tabular}{|c|c|c|}
\hline
$\omega_p/\omega_c$ & $\omega_{DA}/\omega_c = 0.0$ & $\omega_{DA}/\omega_c = 2.0$ \\
\hline
$k_B T / \hbar \omega_c = 1.0$ & 0.961 & 2.266 \\
$k_B T / \hbar \omega_c = 0.5$ & 0.656 & 1.984 \\
$k_B T / \hbar \omega_c = 0.2$ & 0.379 & 1.731 \\
\hline
\end{tabular}
\caption{Optimized values of $\omega_p$ as a function of $T$ and $\omega_{DA}$.}
\label{tab:omega_p}
\end{table}


The following noteworthy observations emerge from a close inspection of Figs. Figs. \ref{fig:ks1wda0}-\ref{fig:ks-3wda2}:
\begin{itemize}
\item In all cases considered, both cavity-free and cavity-modified NE-FGR rate coefficients, $k_{D \rightarrow A} (t)$, reach a plateau following a nonequilibrium transient period, with the plateau value given by the corresponding E-FGR rate constant. This is the expected behavior in condensed-phase systems, where the nonequilibrium initial state equilibrates (often on a time scale comparable to that of CT). Since the effect of the cavity on the E-FGR rate constant was already reported and analyzed in Ref. \citenum{saller_cavity-modified_2023}, we will focus here on the effect of the cavity on the nonequilibrium portion of the dynamics. 
\item Similar to the cavity-free case,\cite{sun_nonequilibrium_2016} the cavity-induced nonequilibrium effects become more pronounced with decreasing temperature ($T$) and decreasing friction ($\eta$) when the molecular system is placed inside the cavity.
\item As for the cavity-free case, the deviation between the cavity modified E-FGR and NE-FGR population relaxation dynamics, $P_D(t)$, extends beyond the initial transient time period during which $k_{D \rightarrow A} (t) \neq  k_{D \rightarrow A}^{eq}$ and increases with decreasing temperature and friction as well as with increasing the shift of the initial state from equilibrium. 
\item Generally speaking, the CT rate is seen to be enhanced by the cavity. The actual enhancement increases with increasing temperature, which can be traced back to the $\langle n_{eq} \rangle$ factor in Eq. \ref{eq:kshifted}. The overall enhancement is also seen to be insensitive to $\eta$. 
\item 
The nonequilibrium effects are seen to be affected by putting the CT system inside the cavity. This can be traced back to the fact that the cavity-modified enhancement of the NE-FGR rates involves $k^m_{j\rightarrow k}(t, \pm \omega_p)$, that correspond to shifting the donor PES by $\pm \hbar \omega_p$, which in turn 
moves the crossing point between the donor and acceptor PESs relative to the initial nonequilibrium configuration of the nuclear DOF. More specifically, the non-equilibrium effects are enhanced (suppressed)  by placing the system in the cavity in cases when shifting the donor PES brings the initial nonequilibrium nuclear configuration nearer to (further away from) the crossing point. Both types of behavior can be seen in Figs. \ref{fig:ks1wda0}-\ref{fig:ks-3wda2}. 
As expected, the cavity-induced nonequilibrium effects are also seen become more pronounced with decreasing temperature and decreasing $\eta$.
\item 
For the cavity-free case in the normal regime, the CT rate in the transient period is generally slower compared to the corresponding E-FGR rate constant. However, the opposite trend is observed in the inverted regime. 
Those trends in the cavity-free NE-FGR rates were previously reported in Ref. 
\citenum{sun_nonequilibrium_2016}, where they were attributed to the fact that  enhancing the nonequilibrium nature of the initial state also enhances its ability to penetrate regions of configuration space with significantly larger overlaps between donor and acceptor nuclear wave functions, which in turn enhances the efficiency of tunneling, thereby giving rise to larger transient CT rates.
In contrast, the cavity-modified rate in the normal regime is often found to be faster compared to the corresponding E-FGR rate constant. This can be attributed to the fact that the cavity-modified enhancement of the NE-FGR rates involves $k^m_{j\rightarrow k}(t, \pm \omega_p)$, which correspond to shifting the donor PES by $\pm \hbar \omega_p$.
Shifting the donor PES up by $\hbar \omega_p$ in particular, pushes the system into the inverted region, thereby leading to a this trend reversal in the cavity modified case. 
\item The ability of IMT to reproduce the NE-FGR rates is seen to increase with increasing temperature and friction, in both the cavity-modified and cavity-free cases. This is to be expected give the fact that IMT is based on short time and high temperature approximations. 
\end{itemize}

%



\section{Summary and concluding remarks}
\label{sec:summary}

FGR-based rate theories have proven to be extremely useful for modeling the dynamics of a wide range of chemical processes. In this paper, we introduced a general-purpose framework for estimating cavity-modified NE-FGR rates of the transition rate between a photoexcited bright donor electronic state and a dark acceptor electronic state, when the
nuclear degrees of freedom start out in a nonequilibrium initial state. Importantly, the proposed framework requires the same inputs needed for estimating the corresponding cavity-free NE-FGR rates. Using this framework therefore makes it possible to bypass the need for an explicit simulation of the molecular system inside the cavity.

The proposed framework is based on the fact that the photonic DOF can be added to the Hamiltonian as nuclear-like DOF, as well as on assuming that the photonic and nuclear DOF are coupled to the electronic DOF, but not to each other. Taking advantage of the resulting separability of the nuclear/photonic terms in the Hamiltonian into purely nuclear and purely photonic terms then makes it possible to calculate cavity-modified NE-FGR rates from cavity-free inputs.

Unlike other cavity-enabled effects,  relatively weak coupling between molecular and cavity DOF can give rise to significant cavity-induced effects on the NE-FGR rates. This is because NE-FGR is already based on treating the coupling between electronic states as a small perturbation (within the framework of second-order perturbation theory). More specifically, coupling between the electronic DOF and cavity modes modifies the coupling between electronic states when the molecular system is placed inside a cavity. Thus, treating this additional electronic coupling as a small perturbation can still give rise to significant modifications of the corresponding NE-FGR rates. 

We also introduced a generalization of IMT for cavity-modified photoinduced CT reactions. IMT corresponds to the shot-time and high-temperature limit of the LSC approximation of NE-FGR with the additional assumption that the fluctuations of the donor-acceptor energy gap satisfy Gaussian statistics. Given the wide range of applicability of those assumptions, cavity modified IMT opens the door to applications to realistic models. 

In addition to significantly modifying the rates, coupling to the cavity was also observed to be capable of affecting the nonequilibrium effects by shifting the dressed donor PES by $\pm \hbar \omega_p$, which in turn modifies the nonequilibrium nature of the initial state.  The cavity-induced nonequilibrium effects were also observed to become more pronounced with decreasing temperature and decreasing friction.

While being rather general, the proposed framework is still based on a number of assumptions, including restricting ourselves to two electronic states and assuming that the molecular system is coupled to a single cavity mode and has no permanent dipole moment. Work on extending the framework beyond those restrictive assumptions, while still maintaining its ability to produce cavity-modified rates from cavity-free inputs, is currently underway and will be reported in forthcoming publications.

\clearpage
\section*{Acknowledgments}
E.G. acknowledges support from the NSF via Grant CHE-2154114.
We thank Prof. Xiang Sun for sharing the parameters for the CPC$_{60}$ harmonic three-state model Hamiltonian.
The top panel of Fig. \ref{fig:system_triad} was created with VMD (VMD was developed by the Theoretical and Computational Biophysics Group in the Beckman Institute for Advanced Science and Technology at the University of Illinois at Urbana-Champaign\cite{VMD}).

\bibliographystyle{aipnum4-1}
\bibliography{bib}

\end{document}